\documentclass[longauth]{aa} 
\RequirePackage{etex}
\usepackage{graphicx}
\usepackage{txfonts}
\usepackage[colorlinks=true, linkcolor=blue, citecolor=blue]{hyperref}
\usepackage{newtxtext,newtxmath}
\usepackage[T1]{fontenc}
\usepackage{amsmath}
\usepackage{amssymb}
\usepackage{color}
\usepackage{pdflscape}
\usepackage{xcolor}
\usepackage{placeins}
\usepackage{subcaption}
\usepackage{orcidlink}
\usepackage{placeins}
\usepackage{longtable}
\usepackage{etoolbox} 

\newcommand{\okina}{'}

\begin{document} 

\title{The type Ia supernova 2023vjh: a peculiar 1991bg-like SN with unusually faint light curves}
\titlerunning{The unusually faint SN~2023vjh}
\authorrunning{Kopsacheili, et al.}
\author{
M. Kopsacheili\inst{1,2}\thanks{E-mail: kopsacheili@ice.csic.es (MK)}\orcidlink{0000-0002-3563-819X},
L. Galbany\inst{1,2}\orcidlink{0000-0002-1296-6887}, 
G. Folatelli\inst{3,4}\orcidlink{0000-0001-5247-1486}, 
M. M. Phillips\inst{5},
C. R. Burns\inst{6}\orcidlink{0000-0003-4625-6629}, 
M. D. Stritzinger\inst{7}\orcidlink{0000-0002-5571-1833},
H.-Y. Miao\inst{1,2}, 
M. González-Bañuelos\inst{1,2}\orcidlink{0009-0006-6238-3598}, 
R. García-Benito\inst{8}\orcidlink{0000-0002-7077-308X}, 
T. E. M\"uller-Bravo\inst{9,10}\orcidlink{0000-0003-3939-7167}, 
E. Y. Hsiao\inst{11}\orcidlink{0000-0003-10392928},
K.~Auchettl\inst{12, 13}\orcidlink{0000-0002-4449-9152},
J. P. Anderson\inst{14}\orcidlink{0000-0003-0227-3451},
C.~Ashall\inst{15}\orcidlink{0000-0002-5221-7557},  
T.-W. Chen\inst{16}\orcidlink{0000-0002-1066-6098}, 
D. D. Desai\inst{15}\orcidlink{0000-0002-2164-859X}, 
M.~E.~Huber\inst{15}\orcidlink{0000-0003-1059-9603}, 
T. de Jaeger\inst{17}\orcidlink{0000-0001-6069-1139}, 
J.~M.~DerKacy\inst{18}\orcidlink{0000-0002-7566-6080}, 
M. Gromadzki\inst{19}\orcidlink{0000-0002-1650-1518}, 
J. T. Hinkle\inst{20,21,15}\orcidlink{0000-0001-9668-2920},  
W. Hoogendam\inst{15}\orcidlink{0000-0003-3953-9532}, 
C. Jimenez-Palau\inst{1,2}\orcidlink{0000-0002-4374-0661},
K. Matilainen\inst{22}\orcidlink{0000-0001-6540-076}, 
P. A. Mazzali\inst{23,24}\orcidlink{0000-0001-6876-8284},
K.~Medler\inst{15}\orcidlink{0000-0001-7186-105X}, 
T. Pessi\inst{14}\orcidlink{0000-0001-6540-0767}, 
C.~Pfeffer\inst{15}\orcidlink{0000-0002-7305-8321}, 
K. Phan\inst{1,2}\orcidlink{0000-0001-6383-860X},
G. Pignata\inst{25}\orcidlink{0000-0003-0006-0188}, 
R. Sanfeliu\inst{1,2}\orcidlink{0009-0007-0178-2875},
B.~J.~Shappee\inst{15}\orcidlink{0000-0003-4631-1149},
M. A. Tucker\inst{26}\orcidlink{0000-0002-2471-8442},
H. Xiao\inst{11}\orcidlink{0009-0006-9436-7197},
D. R. Young\inst{27}\orcidlink{0000–0002–1229–2499}
}

\institute{
Institute of Space Sciences (ICE, CSIC), Barcelona, Spain.\and Institut d’Estudis Espacials de Catalunya (IEEC),  Barcelona, Spain. \and Instituto de Astrofísica de La Plata (IALP), CONICET,  Argentina. 
\and Facultad de Ciencias Astronómicas y Geofísicas (FCAG), Universidad Nacional de La Plata (UNLP), Argentina. 
\and Carnegie Observatories, Las Campanas Observatory, Colina El Pino,  Chile.
\and Carnegie Observatories, Pasadena, USA.
\and Department of Physics and Astronomy, Aarhus University, Denmark.
\and Instituto de Astrofísica de Andalucía–CSIC, Granada, Spain.
\and School of Physics, Trinity College Dublin, The University of Dublin, Dublin, Ireland.
\and Instituto de Ciencias Exactas y Naturales (ICEN), Universidad Arturo Prat Iquique, Chile.
\and Department of Physics, Florida State University, USA.
\and School of Physics, The University of Melbourne, Parkville, VIC, Australia.
\and Department of Astronomy and Astrophysics, University of California, Santa Cruz, CA, USA.
\and European Southern Observatory, Santiago, Chile.
\and Institute for Astronomy, University of Hawai\okina i at Mānoa, Honolulu, USA.
\and Graduate Institute of Astronomy, National Central University, Jhongli, Taiwan.
\and Sorbonne Université, Université Paris Cité, CNRS, Laboratoire de Physique Nucléaire et de Hautes Energies, Paris, France.
\and Space Telescope Science Institute, Baltimore, USA.
\and Astronomical Observatory, University of Warsaw,  Warszawa, Poland.
\and Department of Astronomy, University of Illinois Urbana-Champaign, Urbana, USA.
\and NSF-Simons AI Institute for the Sky (SkAI), Chicago, USA.
\and Department of Physics and Astronomy, University of Turku, Finland.
\and Astrophysics Research Institute, Liverpool John Moores University, Liverpool, UK 
\and Max-Planck-Institut für Astrophysik,  Garching, Germany
\and Instituto de Alta Investigación, Universidad de Tarapacá, Casilla 7D,
Arica, Chile
\and Center for Cosmology and Astroparticle Physics, The Ohio State University, Columbus, USA.
\and Astrophysics Research Centre, Queen’s University Belfast, Belfast, UK
 }
\date{Accepted July 8, 2026}

\abstract{We present  observations of the 1991bg-like type Ia supernova (SN) 2023vjh, associated with the elliptical galaxy MCG+04-10-013. Its light-curve shape parameters ($\Delta m_{15}(B)=1.89 \pm 0.01$ mag and $s_{BV} = 0.45 \pm 0.03$), together with its spectroscopic evolution, place it firmly within the class of fast-declining, subluminous SNe Ia.  The near-peak spectra show prominent features of \ion{Si}{II}, \ion{Ca}{II}, and \ion{Ti}{II}, consistent with a cool photosphere, and place SN~2023vjh within the “cool” and extreme cool regions on the classification diagrams. In addition, the three late-phase near-infrared (NIR) spectra display the \ion{Ca}{II} NIR triplet, \ion{Fe}{II}, and \ion{Co}{II} absorptions, but no obvious $H$-band break. Although SN~2023vjh falls in the same regions of the classification diagrams as other well-studied 91bg-like events, it shows some deviation within this class. In particular, it is systematically fainter than predicted by explosion models and fainter than other well-studied 91bg-like SNe. Light-curve fitting and color-based analyses indicate a relatively large reddening ($E(B-V)_{host}\sim$ 0.2 – 0.35 mag), which is unusual for 91bg-like SNe. Yet, the lack of detectable Na~I~D absorption in its spectra, along with its large projected distance from the center of its passive host galaxy (6.8 kpc), suggests that interstellar extinction along the line of sight is minimal. SN~2023vjh appears to be fainter than typical 91bg-like SNe, and it could be affected by circumstellar material (CSM). Comparisons with explosion models and alternative extinction prescriptions show that including CSM-like extinction improves agreement in the blue bands, but residual discrepancies in the $i$-band may reflect limitations in the model.}


\maketitle

\section{Introduction}
Supernovae are of crucial importance for the interstellar medium (ISM), as they enrich it with heavy elements \citep{1996A&A...315..105R} and deposit large amounts of energy into it. More specifically, Type Ia supernovae (SNe Ia) produce the majority of iron-group elements in the Universe \citep{2003Natur.424..651H}. They are also standardizable candles \citep{1993ApJ...413L.105P}, which enables precise measurements of cosmological parameters such as the Hubble constant (\citealt{1998AJ....116.1009R,1999ApJ...517..565P}).

For normal SNe Ia, there is a tight relationship between peak luminosity and their light-curve decline rate relation, commonly parameterized by $\Delta m_{15}(B)$, the decline in $B$-band magnitude during the first 15 days after maximum light \citep{1977SvAL....3..215P,1993ApJ...413L.105P}, or by the color-stretch parameter $s_{BV}$, which describes the time evolution of the $B-V$ color curve (it is defined as $s_{BV} = t_{max}(B-V)/30$, where $t_{max}(B-V)$ is the maximum of the $B-V$ color curve, measured relative to the epoch of $B$-band maximum; \citealt{2014ApJ...789...32B}), as well as the color at peak (\citealt{1996ApJ...473...88R}; \citealt{1998A&A...331..815T}). These relations make SNe Ia a very useful tool for cosmological distance measurements. Their use led to the unexpected discovery of the accelerated expansion of the Universe and the recognition of dark energy as the dominant component driving it today (\citealt{1998AJ....116.1009R}; \citealt{1999ApJ...517..565P}). SNe Ia continue to provide precise measurements of the expansion rate of the Universe, with the most recent determination of the Hubble constant from normal SNe Ia  $H_0 = 73.18 \pm 0.88\, \rm  km\,s^{-1}\,Mpc^{-1}$ \citep{2025ApJ...992L..34R}.

SNe Ia are found in all types of galaxies, even in non star-forming galaxies, indicating that the progenitors are most probably compact stars. The broad homogeneity of observed properties among SNe Ia suggests a common progenitor system. Although they are widely believed to originate from the thermonuclear explosion of a carbon-oxygen (C-O) white dwarf (WD) in a close binary system \citep{1960ApJ...132..565H}, the specific nature of the progenitor systems and explosion mechanisms remains under debate. Two main scenarios are often considered: the single-degenerate (SD) channel, involving a C-O WD accreting matter from a non-degenerate companion, such as a main-sequence or red giant star \citep{1973ApJ...186.1007W,1990ApJ...354L..53L}, and the double-degenerate (DD) channel, which consists of two C-O WDs (e.g., \citealt{1980SSRv...27..563N}). Observational evidence exists for both. For instance, the absence of a surviving companion in some remnants supports DD origins \citep{2020MNRAS.493.1044T}, whereas events like SN~2012Z, potentially linked to a helium-rich donor (\citealt{2022ApJ...925..138M}; \citealt{2015A&A...573A...2S}; \citealt{2014Natur.512...54M}), and SN~2002ic, associated with a massive AGB star \citep{2003Natur.424..651H}, favor SD systems. Among these progenitor scenarios, four explosion mechanisms are commonly discussed:
(a) Violent mergers of two C–O WDs, where the explosion is triggered during the merging process \citep{2017hsn..book.1237G};
(b) Chandrasekhar-mass explosions, where a WD close to the Chandrasekhar limit ignites centrally due to compressional heating from accretion from the donor star, possibly a red giant, a low-mass (<7–8 $M\odot$) main-sequence star, a He star, or a tidally disrupted WD in a double-degenerate system \citep{1960ApJ...132..565H,1973ApJ...186.1007W,1982ApJ...253..798N,2004MNRAS.353..243P}; 
(c) Sub-Chandrasekhar explosions via the double-detonation mechanism, involving a helium shell detonation that ignites the C–O core \citep{1980SSRv...27..563N,1990ApJ...354L..53L,1994ApJ...423..371W,1996ApJ...457..500H,2017ApJ...846...58H,2010ApJ...719.1067K,2018ApJ...861...78M,2019ApJ...873...84P}; (d) Mergers with masses below, equal, or above the Chandrasekhar mass \citep{1984ApJS...54..335I,1984ApJ...277..355W,2010Natur.463...61P,2013ApJ...770L...8P,2010ApJ...713.1073S,2010ApJ...722L.157V,2013ApJ...778L..18K,2016MNRAS.459.4428K} in triple or quadruple systems, where dynamical interactions induce high-eccentricity orbits leading to a head-on impact \citep{2011ApJ...741...82T,2013MNRAS.435..943P,2013ApJ...766...64S}.

Apart from the normal SNe Ia, there are various subtypes that present different photometric and/or spectroscopic characteristics from the normal SNe Ia, suggesting different explosion mechanisms. One of them is the subluminous subclass of 91bg-like SNe. The prototype of this peculiar class, SN 1991bg, was extensively studied by \citet{1992AJ....104.1543F,1993AJ....105..301L} and \citet{1996MNRAS.283....1T}
Since then, more than 160 91bg-like SNe have been discovered (e.g.  the Transient Name Server
(TNS\footnote{https://www.wis-tns.org/}),  the Carnegie Supernova Project, CSP, \citealt{2006PASP..118....2H,2010AJ....139..519C,2017AJ....154..211K}, \citealt{2022ApJ...928..103H,2024ApJ...960...29P,2026A&A...707A..91A}).

Similar to normal SNe Ia, the spectra of 91bg-like SNe show no evidence of H or He, but they do exhibit strong absorption from Si II \citep{1997ARA&A..35..309F}. Around peak brightness, they also show absorption features from several intermediate-mass elements,  including Si, Mg, Ca, S, and O, which is consistent with their classification as thermonuclear explosions \citep{1997ARA&A..35..309F}.  

However, 91bg-like SNe  present strong  \ion{O}{I} $\rm\lambda$ 7774 and \ion{Ti}{II} lines between 4000 and 4400 \AA\, in the early spectra  (e.g. \citealt{2017hsn..book..317T}, \citealt{2019A&A...630A..76G}). They are characterized by relatively low ejecta velocities, although not dramatically lower than those in normal SNe Ia, with no high-velocity features (HVFs) observed up to now (unlike many SN Ia; \citealt{2005ApJ...623L..37M}). It remains uncertain whether the lack of HVFs in 91bg-like SNe is driven by their explosion physics or by the nature of their surrounding environment (\citealt{2017hsn..book..317T}).

In addition, 91bg-like events usually show notable differences in their light curves  compared to standard SNe Ia, suggesting a different explosion mechanism or progenitor system. Photometrically, 91bg-like SNe are fainter at peak and they rise and decline faster than normal SNe Ia, placing them at the lower right part (i.e. faint and fast-evolving part) of the  width-luminosity relation. The light curve in the near infrared (NIR) bands lacks the secondary maxima which is typical of normal SNe Ia, and hence they present a single peak which occurs a few days later than in B-band (e.g. \citealt{2004ApJ...613.1120G}), while in normal SNe Ia the NIR maxima happens $\sim$ 3 days before than $\rm B_{max}$ (e.g. \citealt{2004ApJ...602L..81K}).

The luminosity-width relation of SNe Ia is assumed to be mainly driven by the amount of $\rm ^{56}Ni$ synthesized during the explosion \citep{1980ApJ...237..541A}. Less luminous SNe are expected to eject smaller amounts of $\rm ^{56}Ni$, typically less than $0.1\,\rm M_\odot$ in the case of 91bg-like SNe \citep{2006A&A...460..793S}, however, the light curve might be affected by the total amount of ejected mass. \citet{2013ApJ...770L...8P} suggest a binary system of He and C-O WDs as the probable progenitor scenario of 91bg-like SNe, since small ejecta mass could explain the narrow light curves. 

Most 91bg-like SNe are found in elliptical galaxies, which are dominated by old stellar populations and lack ongoing star formation \citep{2017hsn..book..317T,2025A&A...694A..10D}. This host galaxy preference supports the idea of older progenitor systems. Only few 91bg-like SNe have been found in spiral galaxies for example SN~1999by, which occurred in the spiral galaxy NGC 2841 \citep{2002ApJ...568..791H, 2004ApJ...613.1120G} and SN~2022xkq in the barred-spiral galaxy NGC 1784 \citep{2024ApJ...960...29P}. Within the Lick Observatory Supernova Search (LOSS) sample \citep{2011MNRAS.412.1441L}, 91bg-like events make up approximately 18\% of all SNe Ia (or higher than 6.7\% according to {\citet{2024MNRAS.530.5016D,2026arXiv260200223D}.

Low-luminosity, fast-declining SN Ia, like the  91bg-like subclass, are often excluded from cosmological analyses because their photometric and spectroscopic properties deviate from those of normal SNe Ia and they do not follow the standard light-curve width–luminosity relation used to standardize SN Ia distances. In particular, their rapid decline rates and intrinsically red colors make them difficult to model with commonly used light-curve fitters, and cosmological samples often remove fast decliners (e.g., objects with $s_{BV}$ <0.5). Nevertheless, several studies have suggested that these events can still be standardized when appropriate parameters are used. Using the color-stretch parameter $s_{BV}$, \citet{2014ApJ...789...32B,2018ApJ...869...56B} showed that fast-declining SNe Ia follow a continuous luminosity–decline relation extending from normal to subluminous objects. \citet{2022ApJ...928..103H} compared the distance moduli of the subluminous SN~2015bo and SN~1997cn with those derived from surface brightness fluctuations \citep{2021ApJS..255...21J} and from $\rm z_{cmb}$-corrected cosmology. They found good agreement and suggested that fast decliners could be as standardizable as normal SNe Ia. More recently, \citet{2024MNRAS.530.4950G} and \citet{2026ApJ...998..101P} revisited the suitability of these objects as distance indicators. In particular, \citet{2026ApJ...998..101P} analyzed a Hubble-flow sample of fast-declining SNe Ia from the Carnegie Supernova Project and demonstrated that distances derived using the Tripp method \citep{1998A&A...331..815T}, where the peak luminosity is modeled as a function of light-curve shape and color, yield values of the Hubble constant consistent with those obtained from normal SNe Ia. They also introduced a simplified “color method,” based solely on the observed ($B_{max}-V_{max}$) pseudo-color at maximum light, which does not require light-curve shape information. Remarkably, this approach provides comparable distance estimates and reflects the strong dependence of peak luminosity on photospheric temperature in fast-declining SNe Ia, rather than being driven purely by dust reddening. These results suggest that, despite being often excluded from cosmological samples, 91bg-like SNe Ia may still serve as useful distance indicators when treated with appropriate standardization methods.

In this paper, we present photometric and spectroscopic observations of SN~2023vjh, a subluminous 91bg-like SN Ia that occurred in the elliptical galaxy MCG+04-10-013. Its photometric and spectral properties are consistent with those of other well-studied 91bg-like events. However, SN 2023vjh appears to be a special case within this subclass, as its dimness and inferred reddening are unusually high compared to typical 91bg-like SNe, despite the lack of clear spectroscopic evidence for significant host-galaxy extinction. The paper is organized as follows: The observations and the data reduction are described in  Sect. \ref{obs_red}. Information on the host galaxy and its properties is presented in Sect. \ref{host}. In Sect. \ref{sec:phot} and Sect. \ref{sec:spec} we further investigate the possible host-galaxy extinction through a series of tests, exploring different empirical and model-based approaches to assess whether the observed faintness and colors can be explained by dust or whether intrinsic effects are required, while also discussing the photometric and spectroscopic properties and their calculation methods. Finally, in Sect. \ref{sum}, we summarize our results.

\section{SN~2023vjh observations and reduction} \label{obs_red}

SN~2023vjh was discovered on 17 October 2023 (MJD = 60234.342) by ZTF with a magnitude of 21.18 in the $g$-ZTF band, and it was subsequently classified as a SN Ia 1991bg-like at $z$ = 0.019 (by the Spectroscopic Classification of Astrophysical Transients group - SCAT; \citealt{2022PASP..134l4502T}). It is possibly associated with the galaxy MCG+04-10-013, an elliptical galaxy at a redshift distance of 83.2 $\pm$ 5.8 Mpc, calculated from the galaxy’s CMB-frame velocity (5642$\pm$8 $\rm km\,s^{-1}$; \citealt{1996ApJ...473..576F}),  a redshift of 0.019 $\pm$ 1.2$\times 10^{-5}$ \citep{2011MNRAS.416.2840L}, and assuming $\rm H_0 = 67.8\, \pm 4.7\, km\, s^{-1}\, Mpc^{-1}$ \citep{2016A&A...594A..13P}. The explosion took place 6.8 kpc (projected distance) from the galaxy center (see Fig. \ref{fig:field}). The main properties of SN~2023vjh and its host galaxy are presented in Table \ref{tab:basic_info}. The photometry sequence and the spectroscopic log are presented in Appendix \ref{sec:logs}

\begin{table}[t]

    \caption{Main properties of the SN~2023vjh and its host galaxy MCG+04-10-013}
    \begin{tabular}{lll}
    \hline
     Parameters    & Values      & Ref  \\
      \hline
      {\bfseries{SN~2023vjh}} & & \\
      \hline
         RA, DEC (J2000) & 04:02:39.2, +25:14:00.5 & (1) \\
             & 60.663250, 25.233533 &(1)  \\
         Explosion date (MJD) & 60234.2 $\pm$ 0.4 & This work \\
         $\rm T_{max}(B)$ (MJD) & 60245.8 $\pm$ 0.4 & This work\\
         $E(B-V)_{MW}$ & 0.236 $\pm$ 0.008 & (2)\\
         $\rm \Delta m_{15}(B) (mag)$  & 1.89 $\pm$ 0.01  & This work \\
         $S_{BV}$ &  0.45 $\pm$ 0.03 & This work\\
     \hline
     {\bfseries{Host galaxy}}& & \\
     \hline
     Name & MCG+04-10-013 & (4) \\
     Type  & -2.7 D & (5) \\
     RA, DEC (J2000)& 04:02:37.9, +25:13:58.3  & (4) \\
     Redshift  & 0.0192 $\pm$ 0.000037 & (6) \\
     Distance (Mpc) & 83.2$\pm$ 5.8 & (6), (7) \\
     $\rm M_B$ & 20.13 $\pm$ 0.53 & (8) \\ 
     \hline
    \end{tabular}
      \small{(1) ATLAS survey \citep{2018PASP..130f4505T}; (2) \citet{2011ApJ...737..103S}; (3)\citet{1998ApJ...500..525S}; (4) Simbad; (5) \cite{2001ApJ...560..566K}; (6) \cite{2011MNRAS.416.2840L}; (7) \citet{1996ApJ...473..576F}; (8) \citet{2003A&A...412...45P}}

\label{tab:basic_info}
\end{table}

\begin{figure}
    \includegraphics[width=0.45\textwidth]{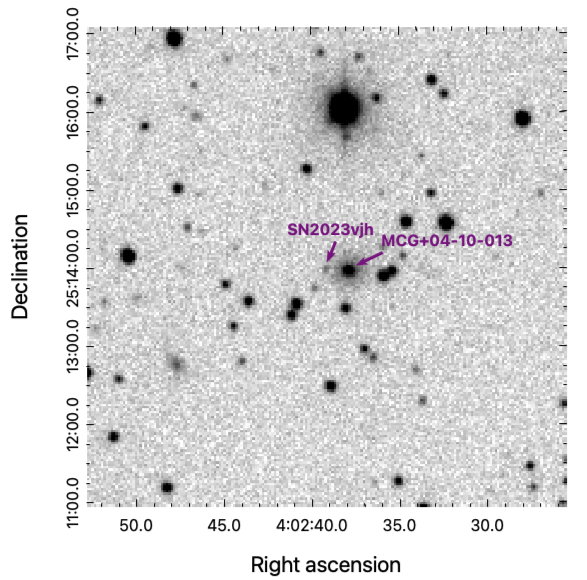}
    \caption{The field of the SN~2023vjh and its host galaxy MCG+04-10-013. The image is taken from the ATLAS RC2 survey (m17 field), retrieved through the DS9 data server interface.}
    \label{fig:field}
\end{figure}

\subsection{Photometry} \label{photometry}
SN~2023vjh was observed as part of the Precision Observations of Infant Supernova Explosions (POISE) program \citep{2021ATel14441....1B}. POISE uses the Henrietta Swope 1 m telescope at Las Campanas Observatory to confirm SN candidates and follow infant SN explosions from public transient survey streams such as the Zwicky Transient Facility (ZTF; \citealt{2019PASP..131a8002B}) and the Asteroid Terrestrial-impact Last Alert System (ATLAS; \citealt{2018PASP..130f4505T}), complemented by spectroscopic observations from multiple facilities.

The POISE follow-up of SN~2023vjh began shortly after discovery and extended over approximately two months.
Optical imaging was obtained in the $B, V, g, r, i, u$ filters and processed using the POISE photometric pipeline, which closely follows the procedures of the Carnegie Supernova Project (CSP; \citealt{2010AJ....139..519C,2017AJ....154..211K}). Briefly, the raw frames are bias- and flat-field corrected using nightly calibration data, and astrometric solutions are derived using the Refcat2 standard star catalog \citep{2018ApJ...867..105T}.

Because POISE is a follow-up program rather than a discovery survey, it does not routinely obtain template images for host-galaxy subtraction. Instead, archival imaging from Pan-STARRS \citep{10.1117/12.457365} or SkyMapper \citep{wolf2018} is used when required.

Photometric calibration is performed by transforming Refcat2 magnitudes of local field stars into the CSP natural $BVgri$ system. The transformation relations were derived from long-term CSP local sequence calibrations and are described in \citet{2017AJ....154..211K} and \citet{2019PASP..131a4001P}.

We also include forced-photometry light curves from ZTF ($g, r$; difference-imaging photometry) and ATLAS ($c, o$; \citealt{2021TNSAN...7....1S}). Detections are defined as observations with a signal-to-noise ratio $>$ 3. The observed magnitudes are presented in Tables \ref{tab:photometry_poise} and \ref{tab:photometry_atlas_ztf} 
for POISE and ZTF/ATLAS, respectively.

 \begin{figure*}[t]
     \centering
     \includegraphics[width=\textwidth]{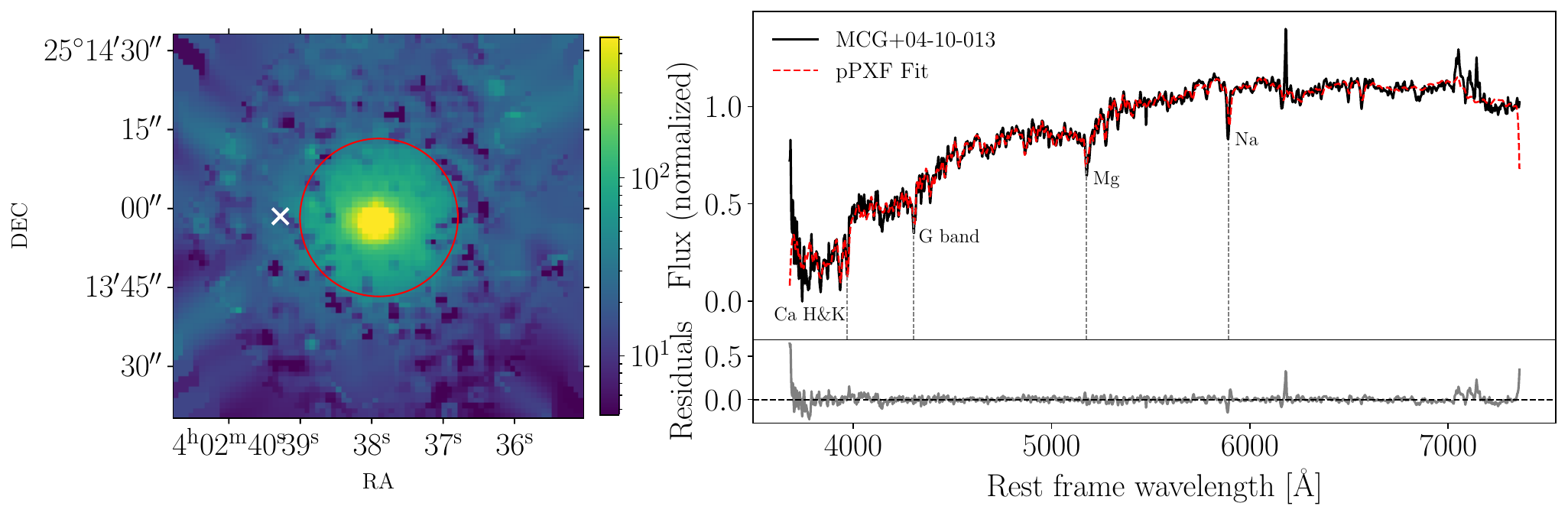}
     \caption{{\it{Left:}} Collapsed 2D image of the galaxy, obtained by summing the flux over all wavelengths (from a PMAS cube at CAHA). The red circle marks the radius containing 90\% of the galaxy light, and the white 'X' indicates the masked position of SN~2023vjh. {\it{Right:}} Spectrum of the galaxy extracted within the same radius (black line) with the best-fit model from pPXF overplotted (red line). The residuals of the fit are shown in gray in the lower panel.}
     \label{fig:galaxy_spec}
 \end{figure*} 
 
For the extinction correction, we adopted the reddening law of \citet{1989ApJ...345..245C},  taking into account the MW  extinction.  The MW component was estimated using  $E(B-V) = 0.236 \pm 0.008$ mag  from \citet{1998ApJ...500..525S,2011ApJ...737..103S},  assuming $R_V = 3.1$.  

\subsection{Spectroscopy} \label{spectroscopy}

\subsubsection{Optical spectra}
The spectra of SN~2023vjh have been obtained with the following telescopes/instruments: University of Hawaii 2.2m telescope (UH88), its Supernova Integrated Field Spectrograph (SNIFS; \citealt{2004SPIE.5249..146L,2022PASP..134l4502T}) instrument, the 10.4m Gran Telescopio CANARIAS (GTC), its Optical System for Imaging and low-resolution Spectroscopy OSIRIS\footnote{https://www.gtc.iac.es/instruments/osiris/} (grisms R1000B and R1000R), the Keck Cosmic Web Imager (KCWI) of the 10m Keck II telescope in Hawaii \citet{2010SPIE.7735E..0MM}, the  Andalusia Faint Object Spectrograph and Camera (ALFOSC) of the 2.56m Nordic Optical Telescope (NOT; \citealt{2010ASSP...14..211D}), the 2.0m Liverpool Telescope (LT; \citealt{2004SPIE.5489..679S}) using the Spectrograph for the Rapid Acquisition
of Transients (SPRAT; \citealt{2014SPIE.9147E..8HP}), the New 3.58m Technology Telescope (NTT) obtained by the  PESSTO collaboration \citep{2015A&A...579A..40S} using the ESO Faint Object Spectrograph and Camera (EFOSC2; \citealt{1984Msngr..38....9B}), the  Inamori-Magellan Areal Camera and Spectrograph (IMACS) of the Magellan Baade Telescope at Chile's Las Campanas Observatory \citet{2011PASP..123..288D}, and the PPAK mode of PMAS \citep{5ecfff171bc740fd9ac42e9f11a9f59f,2006PASP..118..129K} in the 3.5m telescope of the Centro Astronómico Hispano Alemán (CAHA).

Before any kind of analysis, we applied some corrections to the spectra. First, we followed a procedure known as $\it color-matching$ according to which they are scaled to match the photometry (\citealt{2007ApJ...663.1187H}).  Then we applied the MW extinction correction,  and finally we corrected from observed flux to rest-frame flux.

The spectral uncertainties were estimated using a Savitzky–Golay filtering technique. Each spectrum was smoothed with a polynomial fit using window sizes between 41 and 101 resolution elements, chosen to trace the overall continuum while suppressing small-scale features. The difference between the original and smoothed spectra provided residuals that reflect the local noise. The wavelength-dependent root mean square (RMS) of these residuals was adopted as the uncertainty, $\rm \sigma(\lambda)$, which typically increases toward the blue, noisier regions of the spectra. The bluest and reddest portions with flux/$\rm \sigma(\lambda)$ below 3 were excluded from the visual presentation. 
The log of the spectroscopic observations is listed in Table~\ref{tab:spectra}
 
\subsubsection{NIR spectra}

NIR spectra were obtained  at epochs +16.4, +20.8, and +23.4 days with respect to the epoch of $B$ band maximum. Those at +16.4 and +23.4 days were taken with the Folded-port InfraRed Echellette (FIRE) spectrograph installed on the Baade telescope. The data were processed with the IDL {\it{firehose}} pipeline \citep{2013PASP..125..270S}, which is specifically built for reducing FIRE observations. The pipeline carries out flat-field correction, wavelength calibration, sky subtraction, spectral tracing, and spectral extraction. More detailed description of the reduction can be found in \citet{2019PASP..131a4002H}.

The spectrum at epoch +20.8 days was obtained with the Espectr\'ografo Multiobjeto Infra-Rojo (EMIR; \citealt{2022A&A...667A.107G}) mounted on the GTC. Sixteen 120 s exposures were taken using the $YJ$ and $HK$ gratings. The data were reduced with a custom pipeline that performs the standard calibrations, removes telluric absorption using observations of a telluric standard obtained immediately after the SN, and applies the flux calibration. The spectra were corrected for Milky Way  extinction, in the same manner as the optical spectra.

\section{The host galaxy of SN~2023vjh} \label{host}
Integral field spectroscopy (IFS) of the host galaxy MCG+04-10-013 was obtained on 2023 December 16,  with the Potsdam Multi Aperture Spectrograph (PMAS; \citealt{2005PASP..117..620R}) in PPak mode  in the 3.5m telescope of the Centro Astronómico Hispano Alemán (CAHA) at Calar Alto Observatory, as part of the PMAS/PPak Integral-field Supernova hosts COmpilation (PISCO; \citealt{2018ApJ...855..107G}). PPak is a hexagonal bundle and it consists of 331 fibers of 2.7 arcsec each (spatial sampling). The field of view (FoV) of the bundle is $\rm \sim 1.3\, arcmin^2$. 
We acquired three exposures, of 900 s each, with small positional shifts between them. The observations were performed with the V500 grating that covers the  range of 3750 \AA\ - 7500 \AA, with a spectral resolution of 6 \AA. 
 
The optical spectroscopic data  of MCG+04-10-013 were reduced using a customized version of the CAVITY data reduction pipeline \citep{2024A&A...691A.161G}, adapted to the specific requirements of this study. The general workflow follows the standard CAVITY procedure, including CCD frame combination and cleaning, spectral extraction, wavelength and flux calibration, sky subtraction, and final cube reconstruction. The reduction process accounts for Poisson and readout noise propagation, cosmic-ray removal (via PyCosmic; \citealt{2012A&A...545A.137H}), and bias and gain correction for the multi-amplifier CCD configuration used in the PPAK/PMAS instrument. Calibration frames taken within 1.5 h of the science exposures were used to mitigate flexure effects, with wavelength solutions refined using sky lines to ensure subpixel accuracy. Stray light was modeled and subtracted prior to optimal spectral extraction \citep{1986PASP...98..609H}. The extracted spectra were resampled to a common linear wavelength grid, and fiber-to-fiber transmission differences were corrected using twilight flats.

Flux calibration was performed using a sensitivity curve derived from spectrophotometric standard stars observed with the same instrument setup. Atmospheric extinction corrections were applied using airmass-dependent coefficients monitored by the Calar Alto Visual Extinction (CAVEX) system. The initial Galactic extinction correction applied during the PMAS data reduction used the reddening maps of \citet{1998ApJ...500..525S} together with the extinction law of \citet{1989ApJ...345..245C}. However, throughout the remainder of this work, Milky Way reddening corrections are based on \citet{2011ApJ...737..103S}. Therefore, after extracting the supernova spectrum from the PMAS data, we re-applied the Galactic extinction correction using the \citet{2011ApJ...737..103S} reddening value, which corresponds to 0.86 times the $E(B-V)$ from \citet{1998ApJ...500..525S}. As a result, all spectra and photometric measurements presented in this work were corrected consistently using the \citet{2011ApJ...737..103S} extinction estimate. Sky subtraction employed the median of the 30 faintest of the 36 dedicated PPak sky fibers to avoid contamination from field sources.

The final data cube was reconstructed using an inverse-distance weighting interpolation scheme \citep{2012A&A...538A...8S} with a Gaussian kernel of $\sigma$ = 0.75", yielding a spatial sampling of 1.0". Differential atmospheric refraction was corrected iteratively.

We used the IFS data of the galaxy to estimate its age and the stellar metallicity. The first step was to extract the spectrum of the galaxy. To do so, we started by masking out the position of the supernova and replacing the NaN values with interpolated fluxes from the surrounding pixels. In order to define the aperture to extract the spectrum of the galaxy, we collapsed the cube into a 2-dimensional image summing over the spectra, constructed the curve of growth and selected the aperture that contains the 90\% of the total flux (red circle on the left panel of Fig. \ref{fig:galaxy_spec})

The spectrum of MCG+04-10-013 is shown on the right panel of Fig. \ref{fig:galaxy_spec} (black line). It is a typical spectrum of an elliptical galaxy characterized by strong absorption lines due to metals and molecules in the stellar atmospheres. Some of those (Ca H\&K, CH G band, Mg b triplet, Na D doublet) are also indicated in the spectrum.

For the estimation of the stellar age and metallicity we employed the publicly available pPXF (Penalized Pixel-Fitting) code (\citealt{2004PASP..116..138C}, \citealt{Cappellari2023}) to model the observed spectra as linear combinations of simple stellar populations (SSPs) drawn from the MILES stellar library (\citealt{2010MNRAS.404.1639V}, \citealt{2007MNRAS.374..664C}, \citealt{2006MNRAS.371..703S}). The MILES library spans a spectral range of 3525–7500 \AA, with stellar metallicities covering [M/H] from $+$0.22 to $-$2.3, and ages ranging approximately from 0.06 to 18 Gyr. We explored two forms of the Initial Mass Function (IMF): the unimodal (UN) and the bimodal (BI) modes. The unimodal IMF is a single power-law characterized by a logarithmic slope ($\rm \Gamma$), for which  thirteen values ranging from 0.3 to 3.5 are considered. The bimodal IMF shares the same set of slope values, denoted $\rm \Gamma_b$, but differs in shape: it behaves like the unimodal IMF for stellar masses above 0.6 $M_\odot$, while the slope flattens at lower masses, reducing the contribution from low-mass stars through a smooth transition to a shallower profile. The UN mode provided a slightly better fit, and therefore we present the results obtained using this IMF. The output gives the weighted average age and metallicity of the stellar population, with the weights reflecting the relative contributions of each SSP sampled across a regular grid in log(age) and metallicity. The best-fit spectrum, shown in red in Fig. \ref{fig:galaxy_spec}, corresponds to a stellar population with an age of 12.16 Gyr, a metallicity of [M/H] = $-$0.10, and an IMF slope of 0.99 (in this parametrization, a Kroupa/Chabrier-like IMF corresponds to $\Gamma$ = 1.30; \citealt{2001MNRAS.322..231K,2003PASP..115..763C}). Those values are  common in elliptical galaxies that host 91bg-like SNe (e.g., Fig.3 in \citealt{2019PASA...36...31P}).

\FloatBarrier
\begin{figure}
    \centering
    \includegraphics[width=0.5\textwidth]{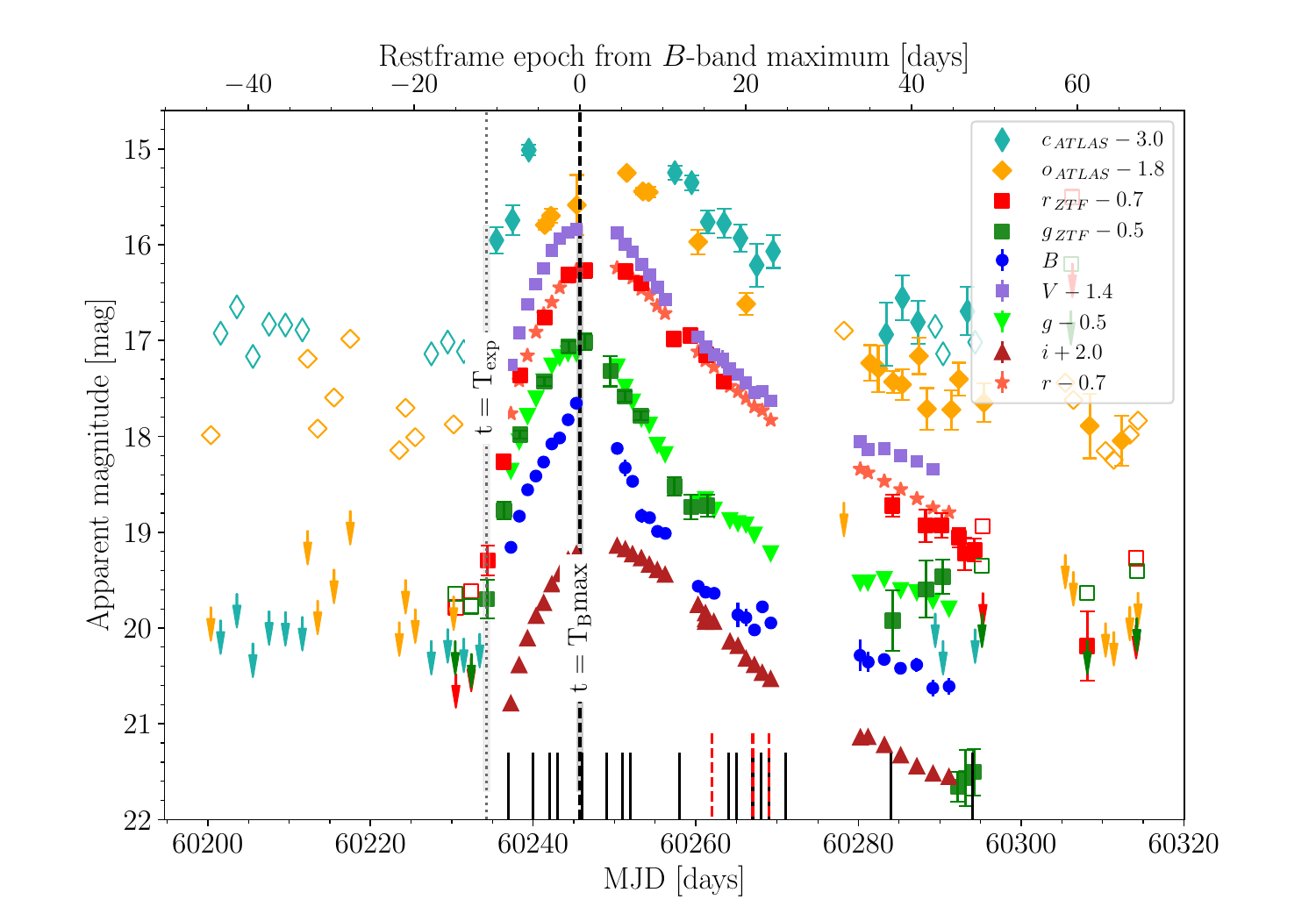}
    \caption{The optical light curves of SN~2023vjh. The gray, dotted line represents the explosion date (see Fig. \ref{fig:exp_time}) and the black, dashed line  the epoch of brightness in the $B$ band obtained from a \texttt{SNooPy} fit (see Fig. \ref{fig:snpy_fit_maxmodel}). The black lines at the bottom represent epochs with optical spectra available, while the red dashed lines those with NIR spectra available.}
    \label{fig:lc}
\end{figure}

\section{Photometric properties}\label{sec:phot}

\subsection{Light-curves}\label{sec:lc}
The light curves {Swope$-BVgri, ZTF-gr$,} and {\it ATLAS-co} of SN~2023vjh are presented in Fig. \ref{fig:lc}. The light curves show a single peak in all bands, with nearly daily coverage from about 11 days before maximum to roughly one month after. We first calculated the explosion time with the most straightforward method, which is the mean value between the last non-detection and the first detection in the {\it ZTF-g} band, and we found $t_0 = 60233.4 \pm 1.0$ MJD.
Then, following a similar approach to \citet{2025JCAP...08..053G}, we performed a simultaneous fit to the early rise of the $g$, $r$, $ZTF-g$, and $ZTF-r$ light curves using a power-law model. The fit included observations from the last two non-detections up to the epoch at which the flux reached half of its maximum value in the red and green bands (POISE and ZTF). All bands were fitted jointly, adopting a single explosion epoch, $t_0$, common to all light curves, while allowing the remaining power-law parameters (e.g., normalization and rise index) to vary independently for each band. For each band, the flux evolution was modeled as:

\begin{equation} \label{eq:exp_date_eq}
f(t) = 
    \begin{cases}
        \begin{array}{lr}
            \alpha(t-t_0)^n+c, \,\,\,t>t_0.
            \hspace{6pt}\\
             c \hspace{1.95cm} t<t_0.  \\
        \end{array}
    \end{cases}
\end{equation}

\noindent The best-fit explosion time is $t_0=60234.19 \pm 0.31 $ MJD (17/10/2023). In Fig. \ref{fig:exp_time} we show the power-law fitting for the $r$ and $g$ bands, along with the best-fit parameters. The power-law indices are $n_r = 1.31 \pm 0.11$ and $n_g = 1.51 \pm 0.17$ for the $r$ and $g$ bands, respectively, in agreement with those of the SN Iax SN~2020udy and SN~2024pxl \citep{2023MNRAS.525.1210M,2025ApJ...988..209H}, and slightly lower than those of the majority of normal SNe Ia ($n_r \sim 2.0$ and $n_g \sim 2.1$; \citealt{2020ApJ...902...47M}).
To determine the date when the light curve reaches its maximum luminosity in the $B$ band ($t_{\mathrm{max}}$) and to derive other properties of the supernova, we used the SuperNovae in object-oriented Python (SNooPy, v2.7.2; \citeauthor{2011AJ....141...19B} \citeyear{2011AJ....141...19B}, \citeyear{2014ApJ...789...32B}) fitter for the $BVgri$,  $r_{ZTF}$, $g_{ZTF}$ bands.
\texttt{SNooPy} fits the light curves for all available filters using light-curve templates trained on the first and second stages of the CSP (\citealt{2017AJ....154..211K}, \citealt{2019PASP..131a4001P}).
It determines the time of maximum light in the $B$ band, the $\Delta m_{15}(B)$ parameter, inferred from all bands, and the magnitude at maximum in each band, when the 'max\_model' is used.
\texttt{SNooPy} also estimates the $s_{BV}$ parameter, which characterizes the color evolution relative to a typical SN~Ia, where the time difference between $B_{\mathrm{max}}$ and $(B-V){\mathrm{max}}$ is 30 days.
 
 For the light-curve fitting, we used a 91bg template with (a) the {\texttt {max\_model}}, which fits the whole light curve in each band independently of the other bands; (b) the {\texttt {EBV\_model2}}, which performs multi-band fits assuming a color law so that all bands are fitted simultaneously; and (c) the {\texttt {color\_model}}, in which intrinsic colors from \citet{2014ApJ...789...32B} are used to infer the amount of extinction $E(B-V)$ and the shape of the reddening law $R_V$. The fits of the three different models are plotted, together with the data, in Fig.~\ref{fig:snpy_fit_maxmodel}. The time of $B$-band maximum is $\mathrm{MJD}=60245.8 \pm 0.4$ for all three models (within uncertainties), corresponding to a rise time of $11.6 \pm 0.5$ days, consistent with the lower end of the rise-time distribution of most SNe~Ia and 91bg-like SNe (e.g. \citealt{2019A&A...630A..76G,2020ApJ...902...47M}).

The color-stretch parameter $s_{BV}$ is $0.448 \pm 0.031$, $0.449 \pm 0.031$, and $0.449 \pm 0.031$ for the {\texttt{max}}, {\texttt{EBV2}}, and {\texttt {color}} models, respectively. The decline-rate parameter $\Delta m_{15}(B)$ is $1.885 \pm 0.005$ mag, as calculated using the {\texttt{max\_model}}. The host-galaxy extinction is $E(B-V)_{host}$ = 0.353$\pm$0.066 
and 0.355$\pm$0.062 mag for the {\texttt{EBV2}} and color models, respectively.
Finally, the shape of the host-galaxy reddening law is $R_v = 2.1 \pm 0.5$, as derived from the {\texttt{color\_model}}. A more detailed discussion of the host-galaxy extinction is presented in Sect. \ref{sec:extinction}.

Although the overall \texttt{SNooPy} template fitting is reasonable, we note that in the $r$-band SN~2023vjh rises more slowly than the template, resulting in a broader light curve, while the $i$-band light curve is narrower than predicted.

\subsection{Extinction from the host galaxy} \label{sec:extinction}
Given that SN~2023vjh is located at a large projected distance of 6.8 kpc from the bulge of its elliptical host galaxy (Fig.~\ref{fig:field}), and that the host is an elliptical galaxy, one would expect little or no extinction (although CSM or a non-negligible line-of-sight contribution cannot be ruled out). However, not all methods or indicators consistently support a scenario of negligible extinction. In the following, we present the different approaches we explored in order to determine whether significant extinction could be present, possibly due to CSM around the progenitor, or whether SN~2023vjh shows intrinsic reddening, which would make it an unusual case.

The first method we used was the {\texttt{SNooPy}} fitting with the {\texttt {EBV\_model2}} and {\texttt{color\_model}}. These models give $E(B-V){\rm _{host}} = 0.353 \pm 0.066$ and $0.355 \pm 0.062$ mag, respectively. Such values are uncommon for 91bg-like SNe, for which the host extinction is typically low, usually $E(B-V){\rm _{host}} < 0.1$ mag, with only a few known exceptions. For instance, in the sample of 14 91bg-like SNe presented by \citet{2024MNRAS.530.4950G}, only one object has a reported host extinction exceeding 0.1 mag ($E(B-V){\rm _{host}} = 0.17 \pm 0.08$ mag), and none of the objects has $E(B-V){\rm _{host}} > 0.2$ mag.

\citet{2014ApJ...789...32B} proposed an updated version of the Lira law \citep{1998AJ....115..234L,2010AJ....139..120F} that is also valid for fast-declining SNe~Ia such as 91bg-like events. Using this updated Lira relation (Eq.~6 in \citealt{2014ApJ...789...32B}) at $t = 45$ days, we obtain $E(B-V)_{\rm host} = 0.20 \pm 0.05$. This value is lower than that inferred from the {\texttt{SNooPy}} fits, but still relatively high given the location of SN~2023vjh within its host galaxy.

On the other hand, the spectra provide no evidence for host-galaxy extinction. In particular, no Na~I~D absorption is detected in any of the spectra of SN~2023vjh. In Fig.~\ref{fig:NaID} we show several high signal-to-noise spectra, where the expected positions of the Na~I~D absorption lines are indicated. No absorption feature associated with the host galaxy is visible, supporting the scenario of negligible extinction along the line of sight within the host (similarly to the 91bg-like SN~2015bo; \citealt{2022ApJ...928..103H}).  We derive a $3\sigma$ upper limit on the equivalent width (EW) of the Na~I~D absorption line from the smoothed spectrum (to avoid variations due to noise) of $<0.53$~\AA\ (or $0.058\pm0.159$~\AA), estimated through Monte Carlo simulations that propagate flux uncertainties and continuum-placement variations.
 
As additional test, we compared the MW-extinction corrected light curves of SN~2023vjh with the models of \citet{2018MNRAS.474.3931B}: DDC25, a Chandrasekhar-mass WD exploding via a delayed deflagration-to-detonation transition, and SCH2p0, a sub-Chandrasekhar-mass WD undergoing a pure central detonation. Finally, we applied the circumstellar dust (CSM) extinction law of \citet{2008ApJ...686L.103G} for both MW- and Large Magellanic Cloud (LMC)-type dust. This CSM extinction law accounts for scattering by dust surrounding the progenitor system, which produces a steeper, low-$R_V$ extinction curve than standard interstellar dust. It is approximated by  $A_\lambda = A_V (1-\alpha +\alpha (\lambda/\lambda_V)^{p})$, with $p = -1.5, -2.5$ and $\alpha = 0.9, 0.8$ for interstellar dust in MW and LMC respectively, leading to stronger attenuation at bluer wavelengths. To test the effect of host extinction, we applied the circumstellar dust extinction law of Goobar (2008) to the light curves after correcting for MW extinction. Assuming $E(B-V)_{\rm host} = 0.2$ we obtain $A_v$ = 0.58, 0.35 for MW and LMC dust models respectively.

In Fig.~\ref{fig:ddc_sch}, we present the light curves of the two models from \citet{2018MNRAS.474.3931B}, DDC25 and SCH2p0, compared to SN~2023vjh under three extinction assumptions: (i) correcting only for MW extinction; (ii) correcting also for host-galaxy extinction with $E(B-V)_{\rm host} = 0.2$ as suggested by the Lira law; and (iii) applying the CSM extinction law of \citet{2008ApJ...686L.103G} only for MW interstellar dust which provides a better match to the data.

The $i$ band remains comparatively fainter than the model predictions even after these corrections. Given that extinction effects are expected to vary smoothly with wavelength, a residual confined primarily to a single red band is less straightforward to attribute solely to reddening uncertainties. This may instead reflect limitations in the model spectral energy distribution in the red/NIR regime, where line-blanketing and Ca~II NIR triplet absorption can have a strong impact on the observed flux.  In addition, we observe a slight discrepancy in the $B$ band as well.

These results suggest that the overall faintness of SN~2023vjh is likely driven by a combination of extinction effects and band-dependent spectral modeling uncertainties. SN~2023vjh is therefore consistent with a fast-declining, subluminous SN~Ia showing possible signs of additional dust or circumstellar material, while the residual tension in the $i$ band highlights the need for improved modeling in the red part of the spectrum. Since it is unclear which  scenario is more likely, we adopt a correction only for MW extinction in the following analysis which is $E(B - V)_{MW}$ = 0.236 $\pm$ 0.008 mag.

\begin{figure}
     \centering
    \includegraphics[width=0.45\textwidth]{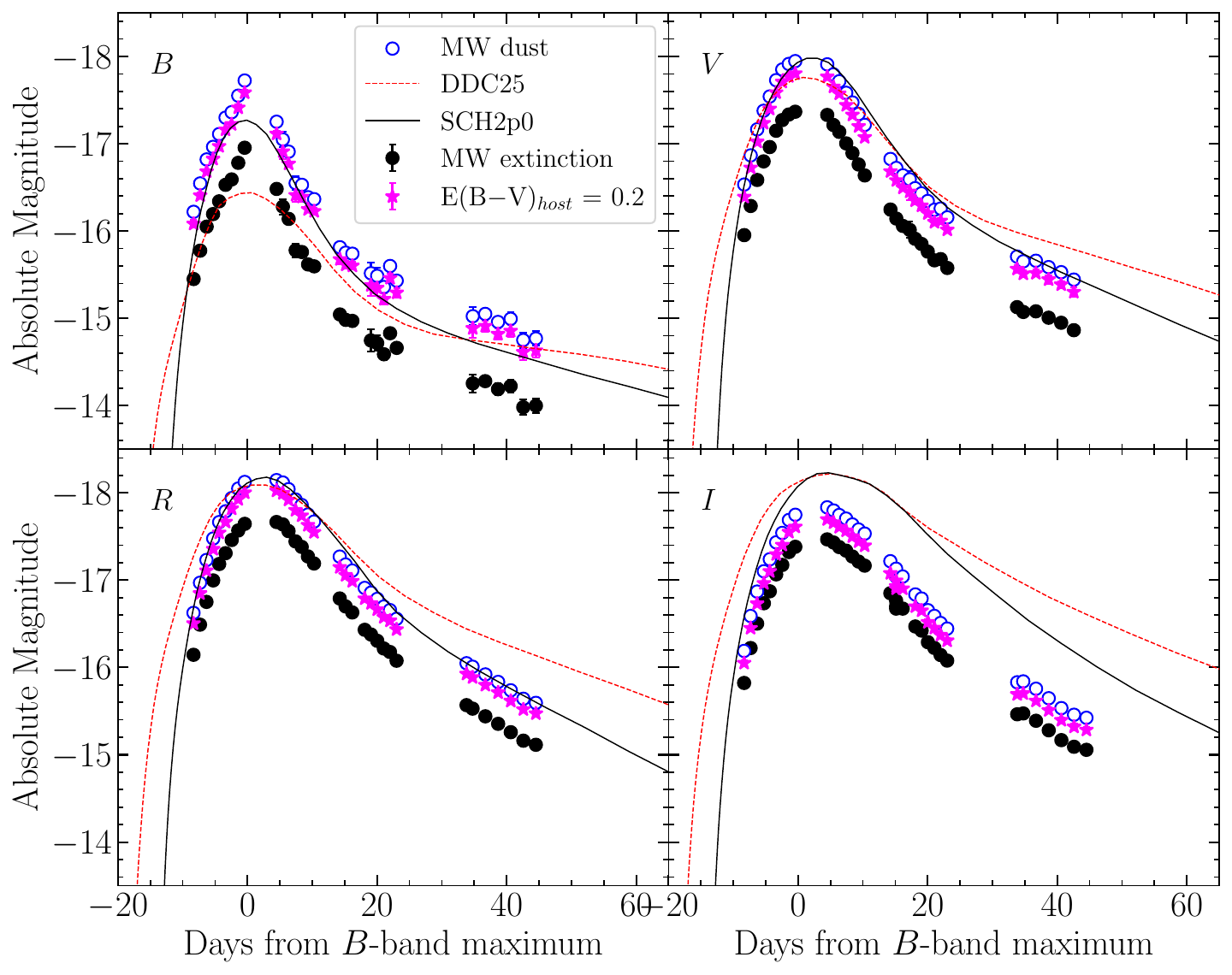}
     \caption{The light curves of the DDC25 and SCH2p0 models from \citet{2018MNRAS.474.3931B}, compared to SN~2023vjh under three extinction assumptions: (i) correcting only for Milky Way extinction; (ii) correcting also for host-galaxy extinction with $E(B-V)_{\rm host} = 0.2$ as suggested by the Lira law \citep{2014ApJ...789...32B}; and (iii) applying the CSM extinction law of \citet{2008ApJ...686L.103G} for Milky-Way interstellar dust. In all cases the light curves are corrected for the Milky Way extinction.} 
     \label{fig:ddc_sch}
 \end{figure}

\subsection{Comparison to light-curves of other SNe Ia} \label{sec:comparison}
Figure \ref{fig:lc_all} shows the light curve of SN~2023vjh compared with those of several other SNe~Ia: the normal SNe~Ia 2011fe \citep{2012JAVSO..40..872R} and 2020jgl \citep{2025JCAP...08..053G}, the transitional SN~1986G \citep{1987PASP...99..592P}, the 2002es-like SN~2002es \citep{2012ApJ...751..142G}, and a small illustrative subsample of 91bg-like SNe: 1991bg \citep{1992AJ....104.1543F, 1993AJ....105..301L, 1996MNRAS.283....1T, 2004AJ....128.3034K}, 1999by \citep{2010ApJS..190..418G, 2004ApJ...613.1120G}, 2005bl \citep{2008MNRAS.385...75T, 2015ApJS..220....9F, 2009ApJ...700..331H, 2010AJ....139..519C}, 2005ke \citep{2015ApJS..220....9F, 2009ApJ...700..331H, 2010AJ....139..519C, 2017AJ....154..211K}, 2006bd \citep{2006CBET..477....1M}, 2006mr \citep{2010AJ....139..519C, 2011AJ....142..156S}, 2007N \citep{2007CBET..818....1L},  2021qvv \citep{2023MNRAS.526.2977G}, and 2022xkq \citep{2024ApJ...960...29P}. In later sections, we will consider these and additional 91bg-like SNe Ia for a more comprehensive comparison with SN~2023vjh. Overall, the light curve of SN~2023vjh is generally consistent with those of the other 91bg-like SNe. It appears to evolve more slowly than SNe 2006mr, 2005ke, 1999by, and 1991bg, while it more closely follows the light curves of 2005bl, 2007N, and 2006bd.

In Fig. \ref{fig:LWR}, the left panel shows the $B$-band absolute magnitude as a function of $\Delta m_{15}(B)$ for SNe Ia from the CSP-I sample \citep{2017AJ....154..211K} and the sample of 91bg-like SNe Ia from \citet{2024MNRAS.530.4950G}. The lower-right region of the diagram is populated by the subluminous, fast-declining 91bg-like SNe Ia, where SN~2023vjh is also located. We show the location of SN~2023vjh assuming only MW extinction correction as a red filled star, and as a red open star when CSM extinction from \citet{2008ApJ...686L.103G} for MW-type dust is included. The right panel presents the $B$-band absolute magnitude as a function of the color-stretch parameter $s_{BV}$. In both diagrams, and under both extinction assumptions, SN~2023vjh lies well within the locus of 91bg-like SNe Ia.

\begin{figure}
    \centering
    \includegraphics[width=0.45\textwidth]{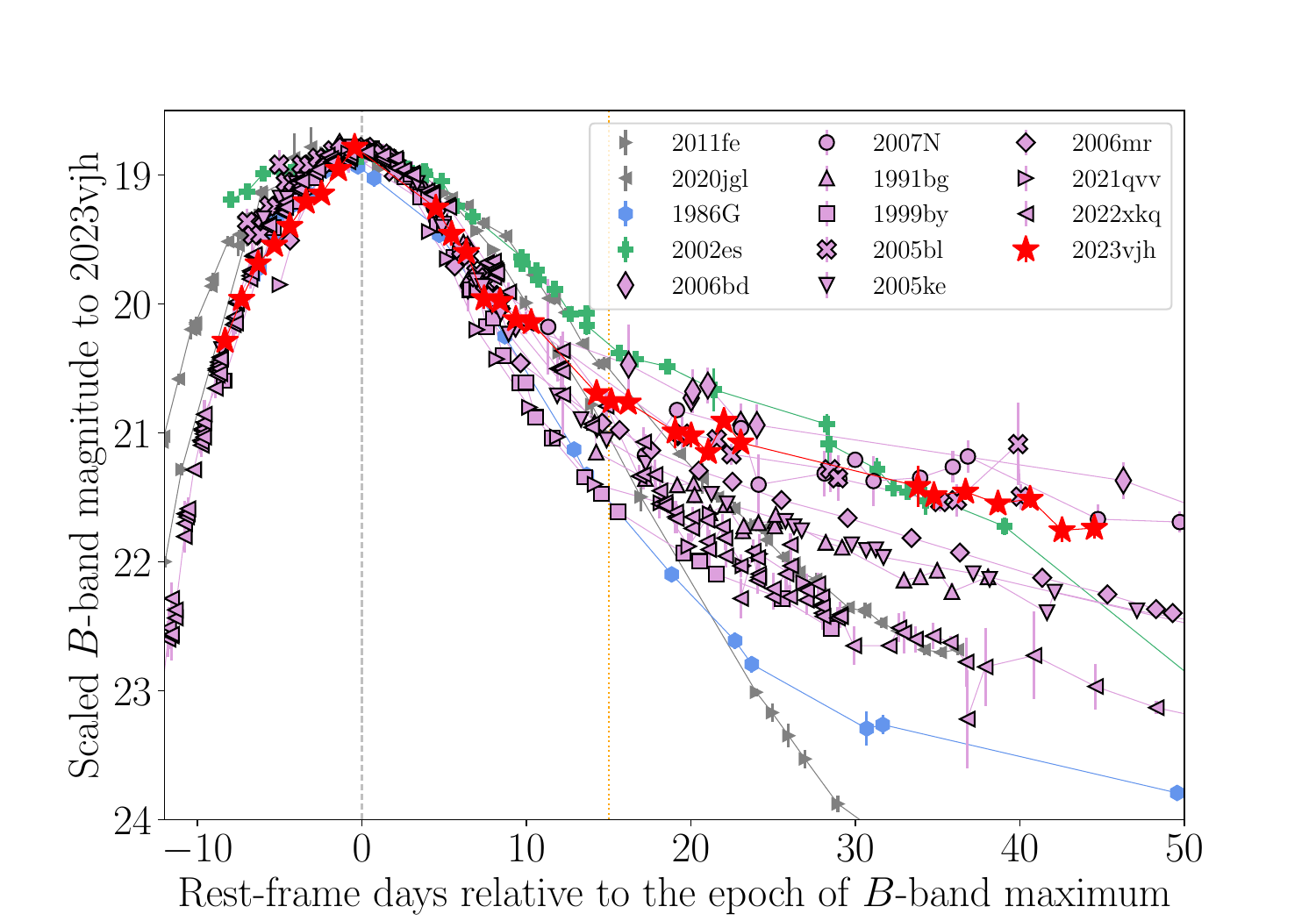}
    \caption{The light-curve of SN~2023vjh (red line) in comparison with light-curves of other 91bg-like SNe (SN~2006bd, SN~2007N, SN~1991bg, SN~1999by, SN~2005bl, SN~2005ke, SN~2006mr, SN~2021qvv, SN~2022xkq, shown in purple), the normal SNe Ia SN~2011fe and SN~2020jgl shown in gray, the transitional SN~1986G in blue, and the SN~2002es in green. The  gray, vertical, dashed-line indicate is at zero where we consider the date of the maximum light, while the orange, dotted line indicates the epoch +15 (references for the SNe are given in Sect.~\ref{sec:comparison}).}
    \label{fig:lc_all}
\end{figure}

\begin{figure}
    \centering
    \includegraphics[width=0.5\textwidth]{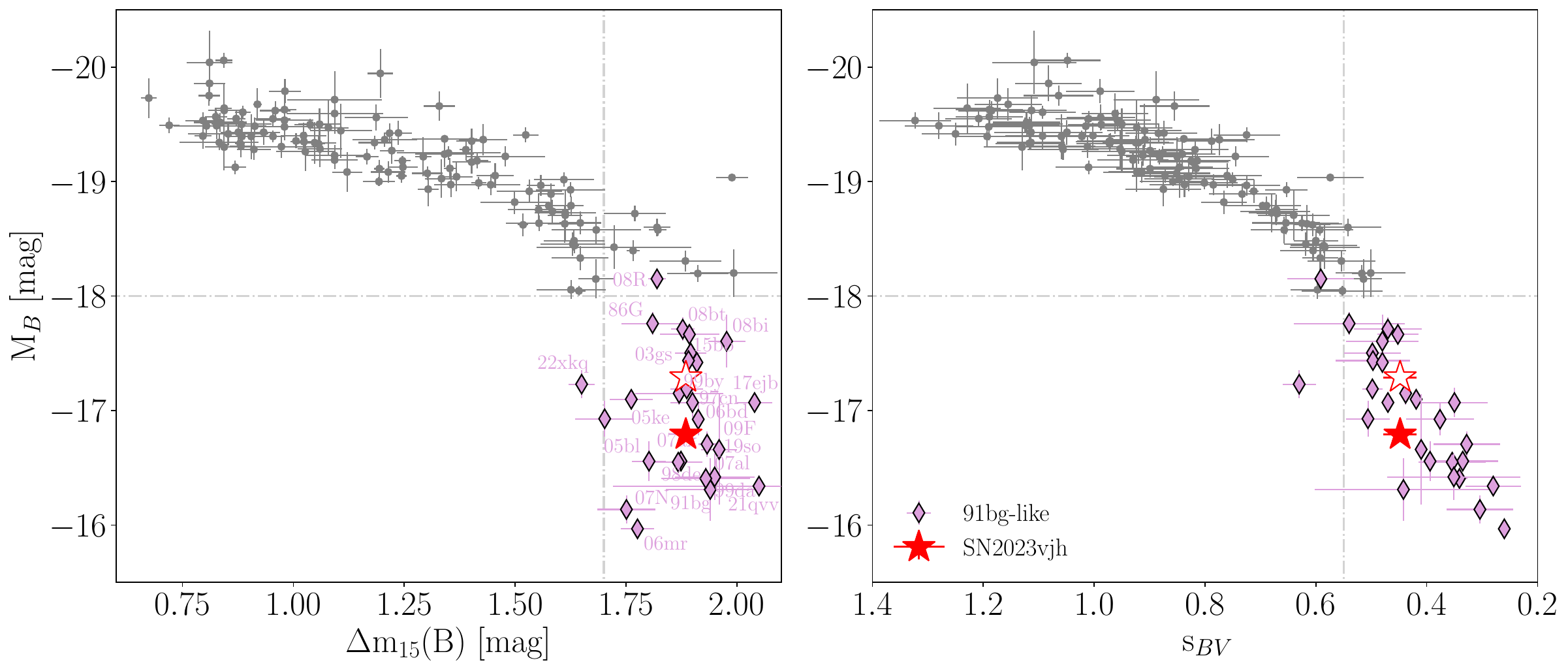}
    \caption{Luminosity width relation as a function of $\Delta$m$_{15}$(B) (Left) and
s$_{BV}$ (right) for a selection of CSP-I SNe Ia from \citet{2017AJ....154..211K} and the sample of \citet{2024MNRAS.530.4950G} with the addition of SN~2023vjh taking into account only the MW extinction correction (red filled star), and assuming an CSM extinction from \citet{2008ApJ...686L.103G} (red open star). The purple diamonds represent 91bg-like SNe.}
    \label{fig:LWR}
\end{figure}

\subsection{Color-curves}\label{sec:cc}
Figure \ref{fig:colors} shows the MW extinction corrected $B-V$, $g-r$, and $r-i$ color curves of SN~2023vjh (top). The bottom panel presents the $B-V$ color curve of SN~2023vjh corrected for MW extinction, compared with other SNe that are also corrected only for MW extinction (91bg-like SNe Ia, a 2002es-like SN, and a normal SN~Ia).
Although our color coverage does not extend far in time (up to $\sim$40 days after maximum), the evolution of SN~2023vjh closely follows that of the other 91bg-like SNe, particularly between $B_{\mathrm{max}}$ and $(B-V)_{\mathrm{max}}$. Overall, it tends to display slightly bluer ($B-V$) colors than most objects in this class.  This may also be related to the slight discrepancy between the observed $B$-band light curve and the model predictions shown in Fig.~\ref{fig:ddc_sch}.

Taking advantage of the high cadence of the observations, we calculated the color-stretch parameter $s_{BV}$ from the $B-V$ color curve. We determined the $(B-V)_{max}$ by fitting a smoothed cubic spline to the color-curve. A smoothing factor of 25 was adopted, after testing various values, to suppress small-scale fluctuations, while preserving the global color evolution. The epoch of maximum color was defined as the time at which the fitted spline reaches its maximum value and the color-stretch parameter was computed as $s_{BV} = t_{max}(B-V)/30$ \citep{2014ApJ...789...32B}. Uncertainties on $s_{BV}$ were estimated via Monte Carlo resampling. The $B-V$ measurements were perturbed according to their uncertainties, the spline fitting and peak determination were repeated for 500 realizations using the same smoothing factor, and the standard deviation of the resulting $s_{BV}$ distribution was adopted as the uncertainty. We found $s_{BV} = 0.45 \pm 0.17$, in excellent agreement with the value derived from {\texttt{SNooPy}}'s best-fit parameters.

\begin{figure}
    \centering
    \begin{subfigure}{\columnwidth}
        \centering
        \includegraphics[width=\columnwidth]{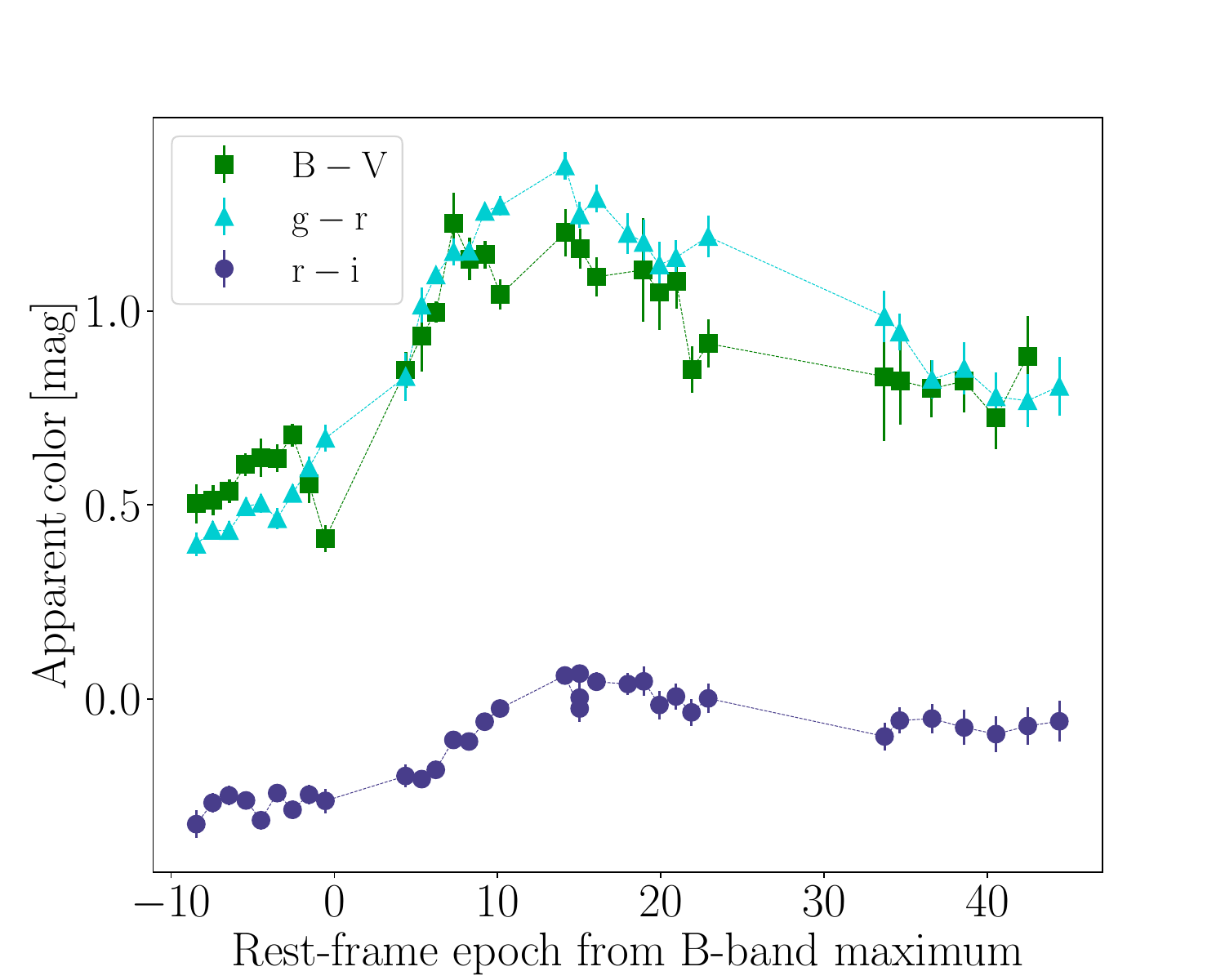}
    \end{subfigure}

    \vspace{0.4em}

    \begin{subfigure}{\columnwidth}
        \includegraphics[width=0.92\columnwidth]{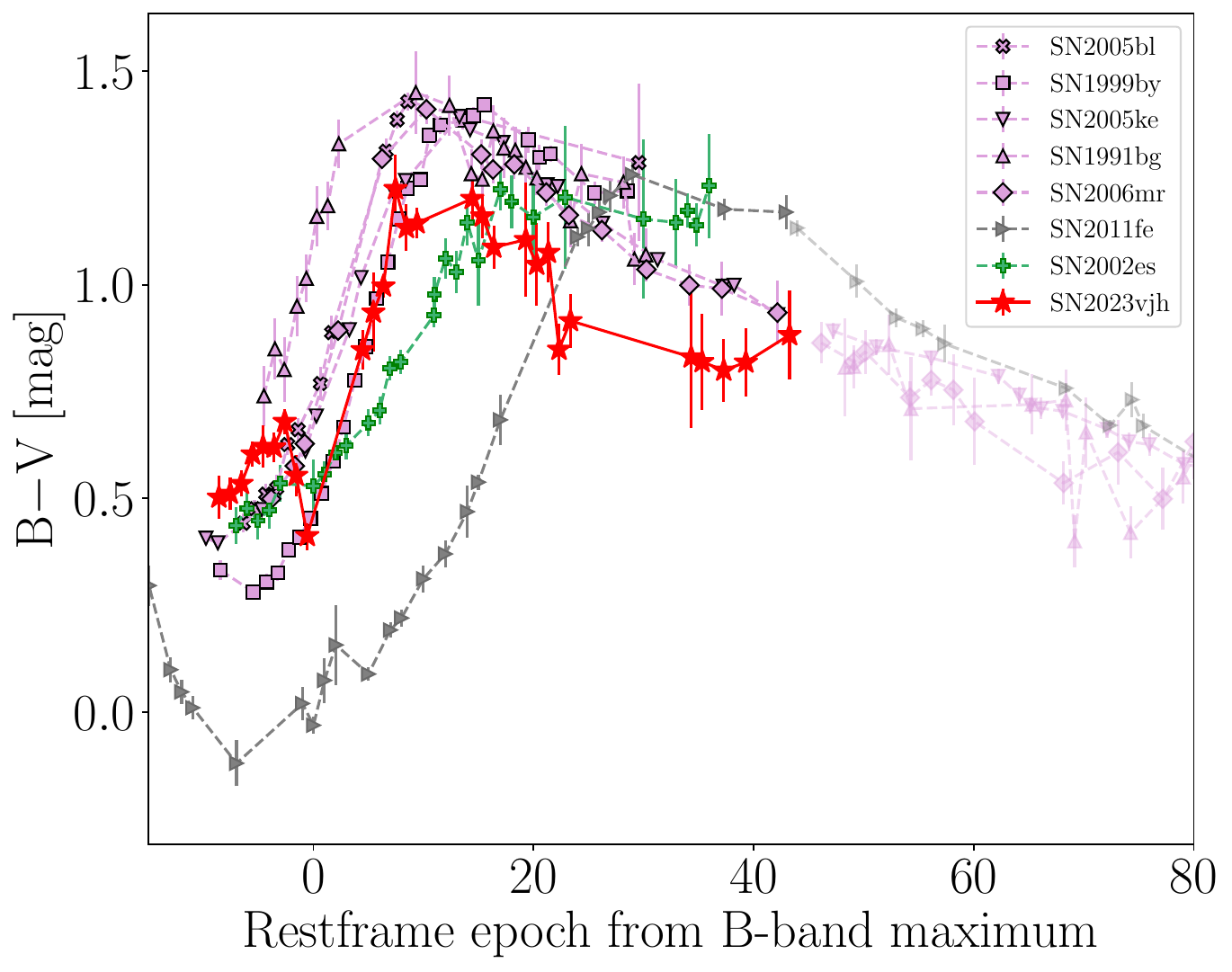}

    \end{subfigure}

    \caption{The MW extinction corrected color curves of SN~2023vjh (top panel) and the MW extinction corrected $B-V$ color curve of SN~2023vjh in comparison  with those of the 91bg-like sample, the normal SN Ia SN~2011fe, and the SN~2002es (bottom panel).}
    \label{fig:colors}
\end{figure}

Finally, in Fig. \ref{fig:SBV_tiBmax} we present the color stretch parameter $s_{BV}$ as a function of $t_{max}^i - t_{max}^B$ for various sub-types of SNe Ia. This method was suggested by \citet{2020ApJ...895L...3A} to photometrically differentiate between various sub-types of. SN~2023vjh lies well within the 91bg-like region, which is generally characterized by lower $s_{BV}$ values compared to other subtypes. 

\subsection{Hubble constant estimate}
We also estimate the value of the Hubble constant using SN~2023vjh, following the approach presented by \citet{2026ApJ...998..101P}, which was specifically developed for fast-declining SNe~Ia. In that work, two methods are proposed to standardize the luminosities of such objects: the classical Tripp relation \citep{1998A&A...331..815T}, which combines light-curve shape and color information, and the Color method, which uses the color at maximum light as a luminosity indicator. In these methods, the distance moduli are computed following Eqs. 1 and 5 of \citet{2026ApJ...998..101P}:
\begin{equation}
    \mu_{Tripp} = m_x -P^{N}_x(s_{BV - 0.5})-\beta_{x}(B_{max}-V_{max}-0.4)
\end{equation}
and 
\begin{equation}
    \mu_{color} = m_x -P^{N}_x(B_{max}-V_{max}-0.4)
\end{equation}

\noindent{where $m_x$ is the peak apparent magnitude in filter $x$, $P^{N}_x$ denotes an N-th order polynomial (we adopt N=2) and $\beta_{x}$  is the slope of the luminosity-color relation). The coefficients of $P^{N}_x$ and the values of $\beta_{x}$ are taken from Table 4 of \citet{2026ApJ...998..101P}. }

Using these calibrations, we compute the value of $H_0$ in each available photometric band and derive weighted averages for both methods. The resulting values are $H_0 = 71.56 \pm 3.09$ km s$^{-1}$ Mpc$^{-1}$ from the Tripp method and $H_0 = 74.60 \pm 2.22$ km s$^{-1}$ Mpc$^{-1}$ from the Color method. Figure~\ref{fig:H0_color_trip} shows the values obtained in each filter together with the weighted averages and the ranges reported by \citet{2026ApJ...998..101P}.

Both estimates are consistent within the uncertainties with the values reported by \citet{2026ApJ...998..101P}. The Color-method result lies somewhat closer to their measurement, while the Tripp-based estimate falls toward the lower end of the reported range. We note that the $i$-band systematically yields lower values of $H_0$, consistent with the fainter $i$-band light curve of SN~2023vjh compared to the templates.
\begin{figure}
    \centering
    \includegraphics[width=0.44\textwidth]{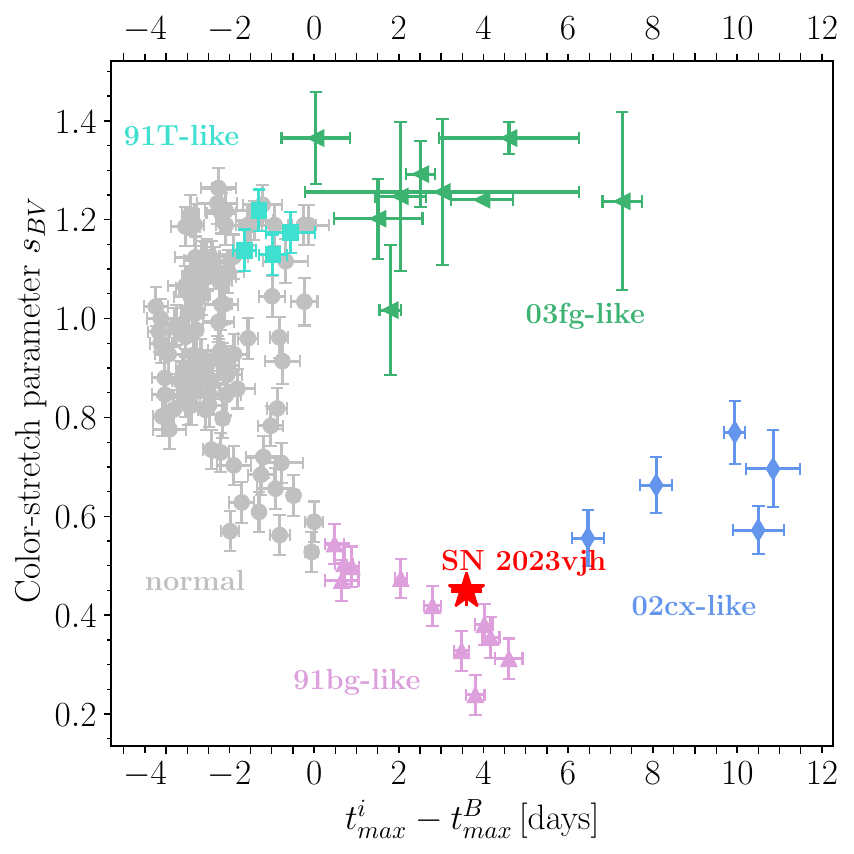}
    \caption{Color-stretch parameter $s_{BV}$ as a function of $t_{max}^i - t_{max}^B$ for different sub-types of SN Ia \citep{2020ApJ...895L...3A}. The $t_{\rm max}^{i}$ and $t_{\rm max}^{B}$ refer to the epochs of maximum light in the $i$ and $B$ bands, respectively.}
    \label{fig:SBV_tiBmax}
\end{figure}

\begin{figure}
    \includegraphics[width=0.45\textwidth]{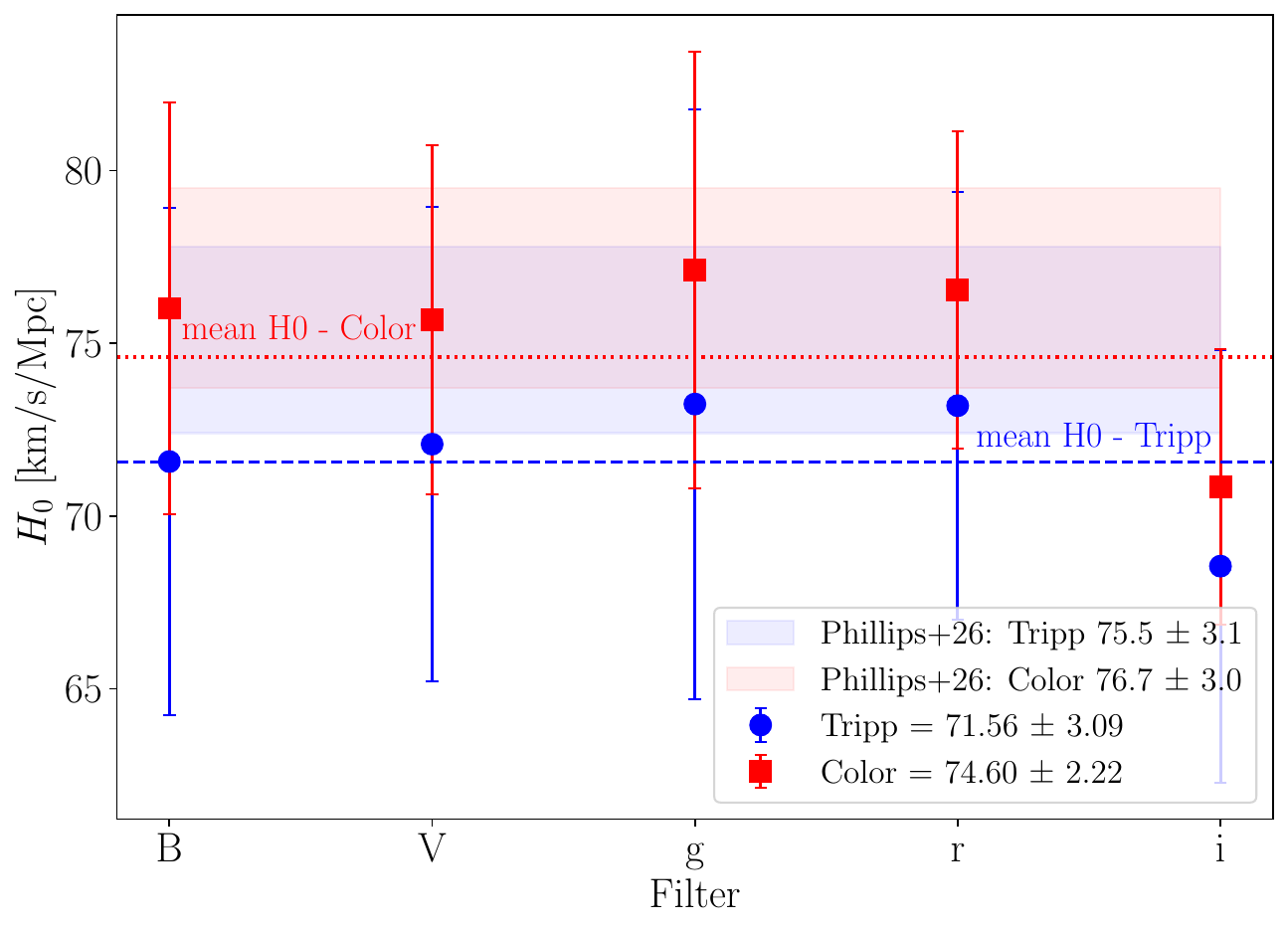}
    \caption{We derive the Hubble constant from SN 2023vjh using both the Tripp method \citep{1998A&A...331..815T} and the Color method \citep{2026ApJ...998..101P}. For comparison, we also include the values reported by \citet{2026ApJ...998..101P} for fast-declining SNe Ia.}
    \label{fig:H0_color_trip}
\end{figure}

\section{Spectroscopic properties}\label{sec:spec}
In this section, we describe the observed spectroscopic properties of SN~2023vjh and compare them with those of other 91bg-like SNe. The rest-frame optical spectra, corrected for MW extinction, are presented in Fig.~\ref{fig:my_spectra}. The original spectra are shown in gray, the smoothed spectra in black.

The available spectra cover phases from approximately $-9$ to $+47$~days relative to the $B$-band maximum. The spectra exhibit prominent absorption features due to \ion{Ca}{II}\,H\&K, \ion{Si}{II} $\lambda\lambda5972,6355$, \ion{O}{I}\,$\lambda7775$, and \ion{Ca}{II}\,IR triplet, which become visible around $-$6 day. The \ion{Si}{II} $\lambda5972$ feature disappears around day~$+6$, while the remaining lines gradually fade by day~$+19$.

\begin{figure}
    \centering
    \includegraphics[width=0.5\textwidth]{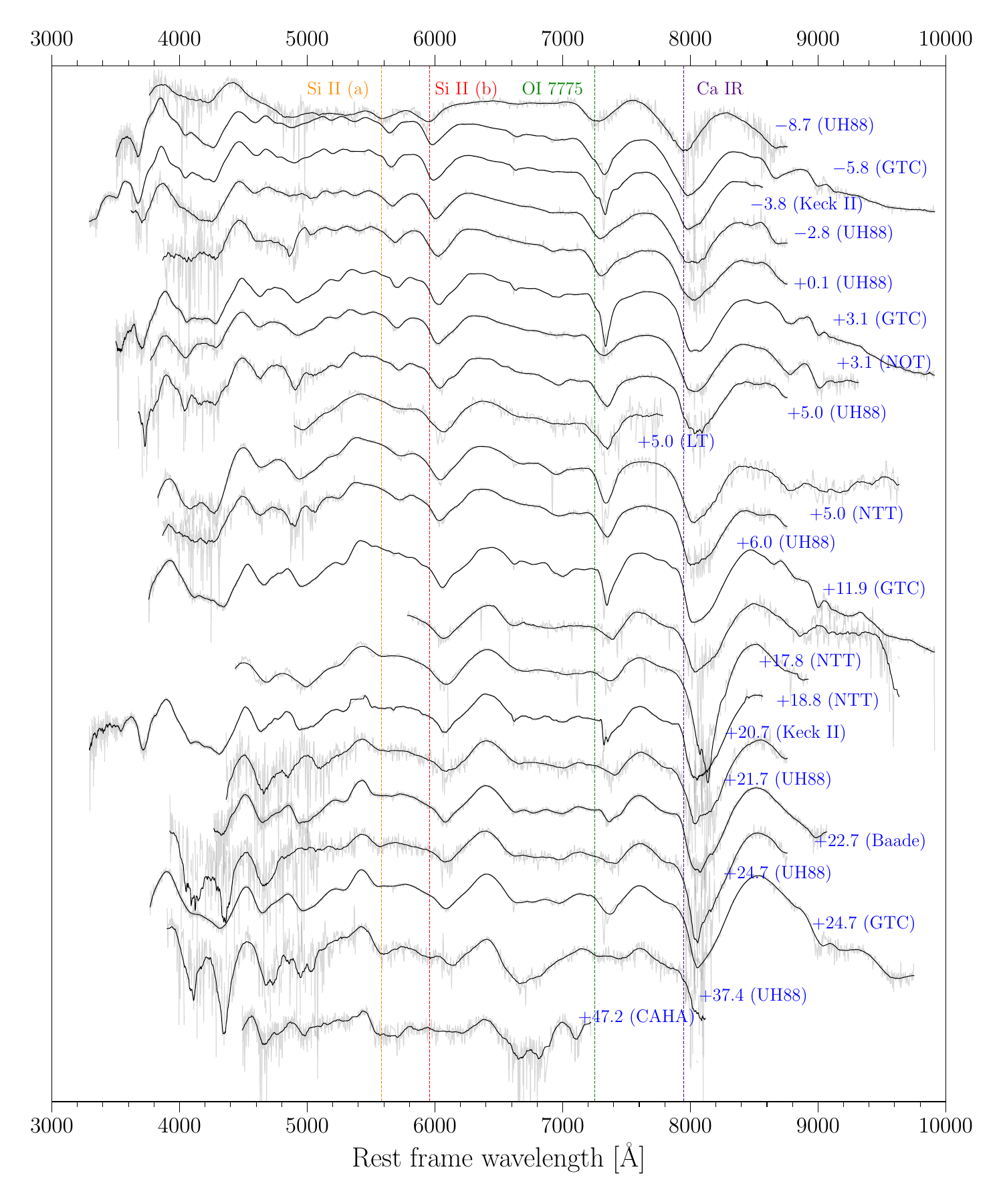}
    \caption{Optical spectra of SN~2023vjh. Here there are 21 observations for 17 different dates.  The original, rest-frame spectra, corrected for color and Milky Way extinction, are shown in gray, while the smoothed spectra are shown in black. The blue labels next to each spectrum denote the telescope and phase relative to B maximum.}
    \label{fig:my_spectra}
\end{figure}

In Fig.~\ref{fig:spectra_comp}, we compare the spectrum around
$B$-band maximum of SN~2023vjh with those of SN~1991bg \citep{1992AJ....104.1543F}, the transitional SN~1986G \citep{1987PASP...99..592P}, and the normal SN~2011fe \citep{2011Natur.480..344N}. SN~2023vjh shows spectral features very similar to SN~1991bg, whereas, compared to the other two SNe, we notice differences in the 4000–4300 \AA\ range, where 91bg-like SNe are dominated by deep \ion{Ti}{II} $\sim$ 4150 \AA\ feature and consequently, SN~1986G and SN~2011fe show lower equivalent widths. In addition, the \ion{Si}{II} $\lambda5972$ absorption line is stronger in SN~2023vjh and SN~1991bg compared to the others, as first reported for fast-declining (low-luminosity) SNe by \citet{1995ApJ...455L.147N} (also \citealt{2008MNRAS.389.1087H}).

\begin{figure}
    \centering
    \includegraphics[width=0.45\textwidth]{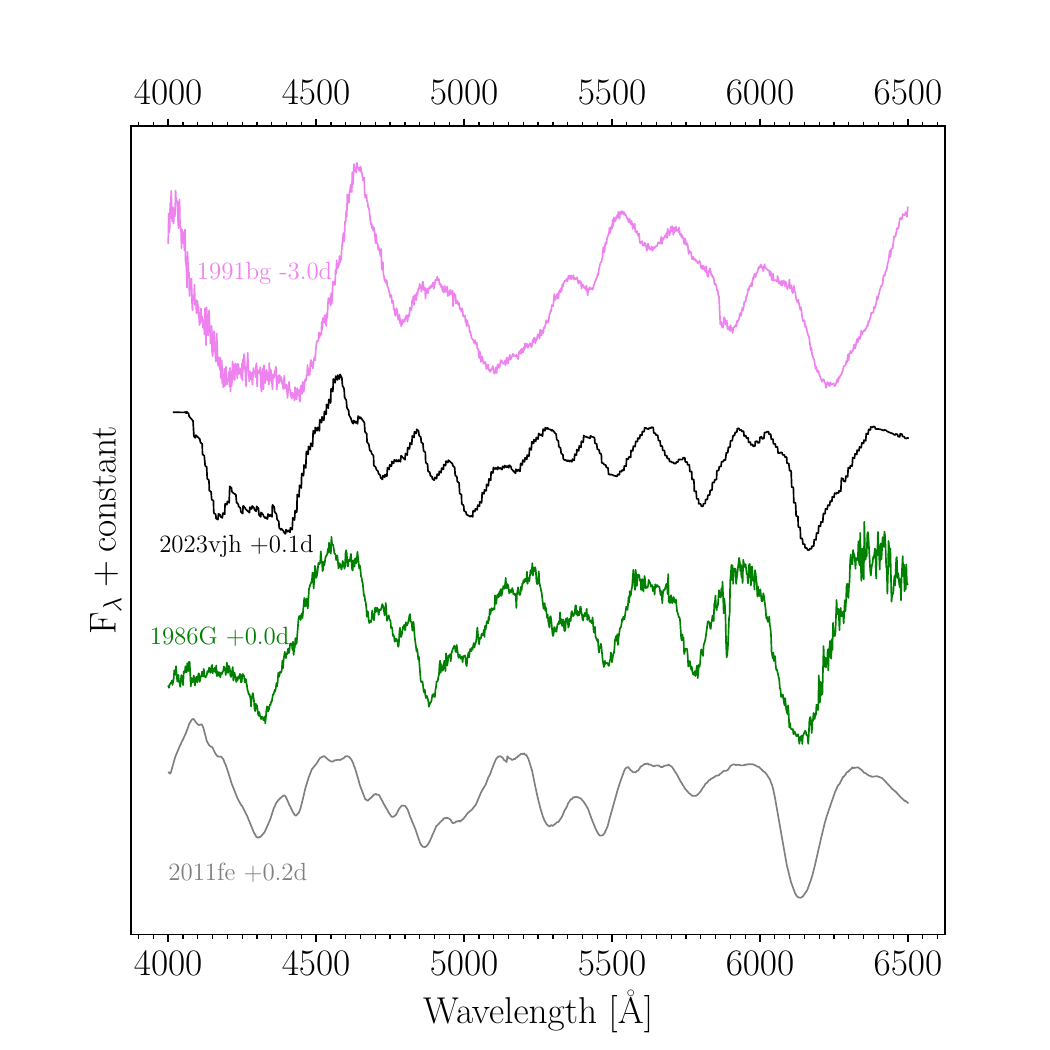}
    \caption{The optical spectrum of SN~2023vjh obtained around the epoch of $B$-band maximum compared with similar epoch spectra of SN~1991bg, the transitional SN~1986G, and the normal SN~2011fe.}
    \label{fig:spectra_comp}
\end{figure}

\begin{figure}
    \centering
    \begin{minipage}{0.45\textwidth}
        \centering
       \includegraphics[width=\textwidth]{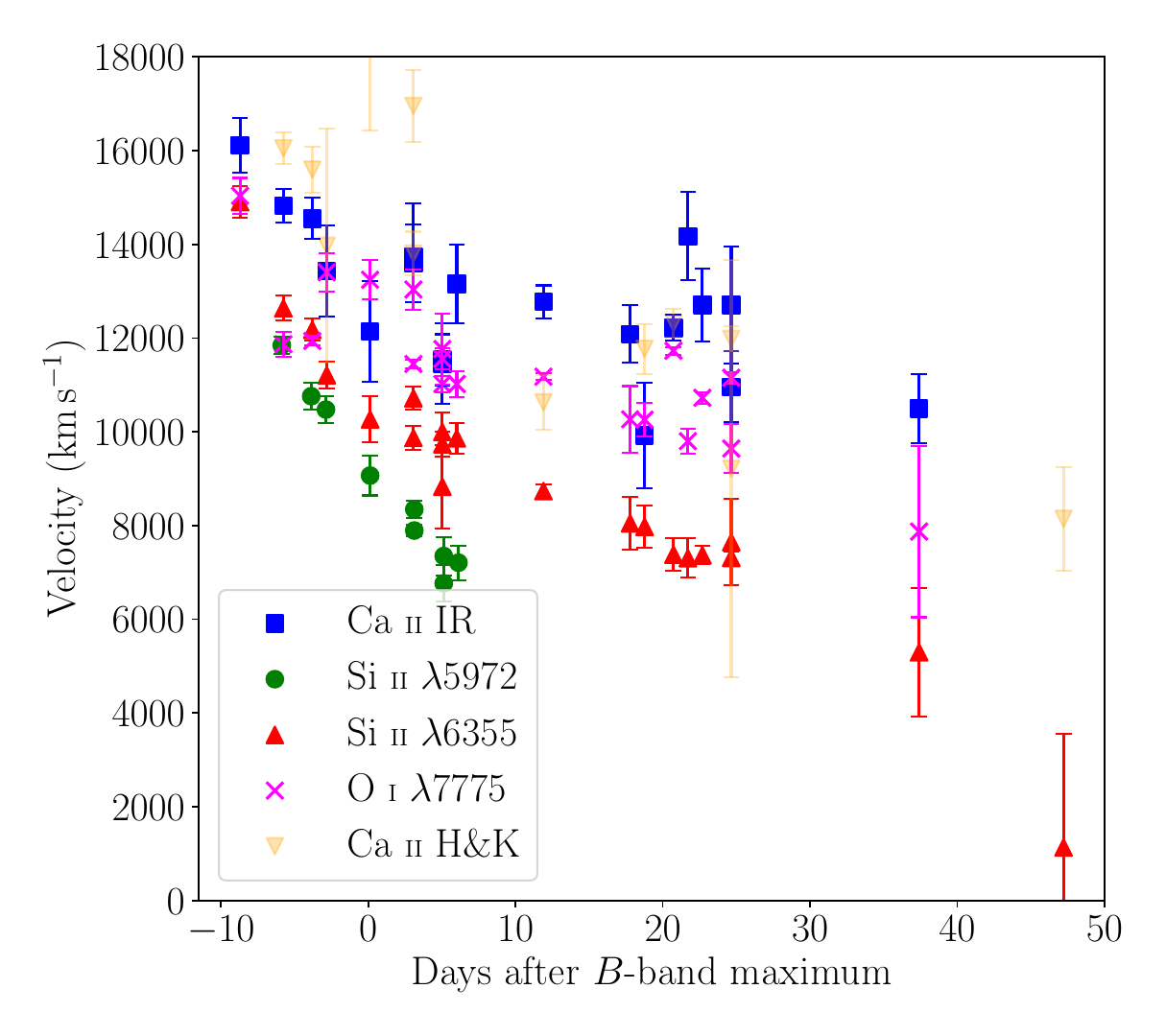}
\caption{Evolution of the expansion velocities.}
    \label{fig:velocities}
    \end{minipage}
    \hfill
    \begin{minipage}{0.45\textwidth}
        \centering
        \includegraphics[width=\textwidth]{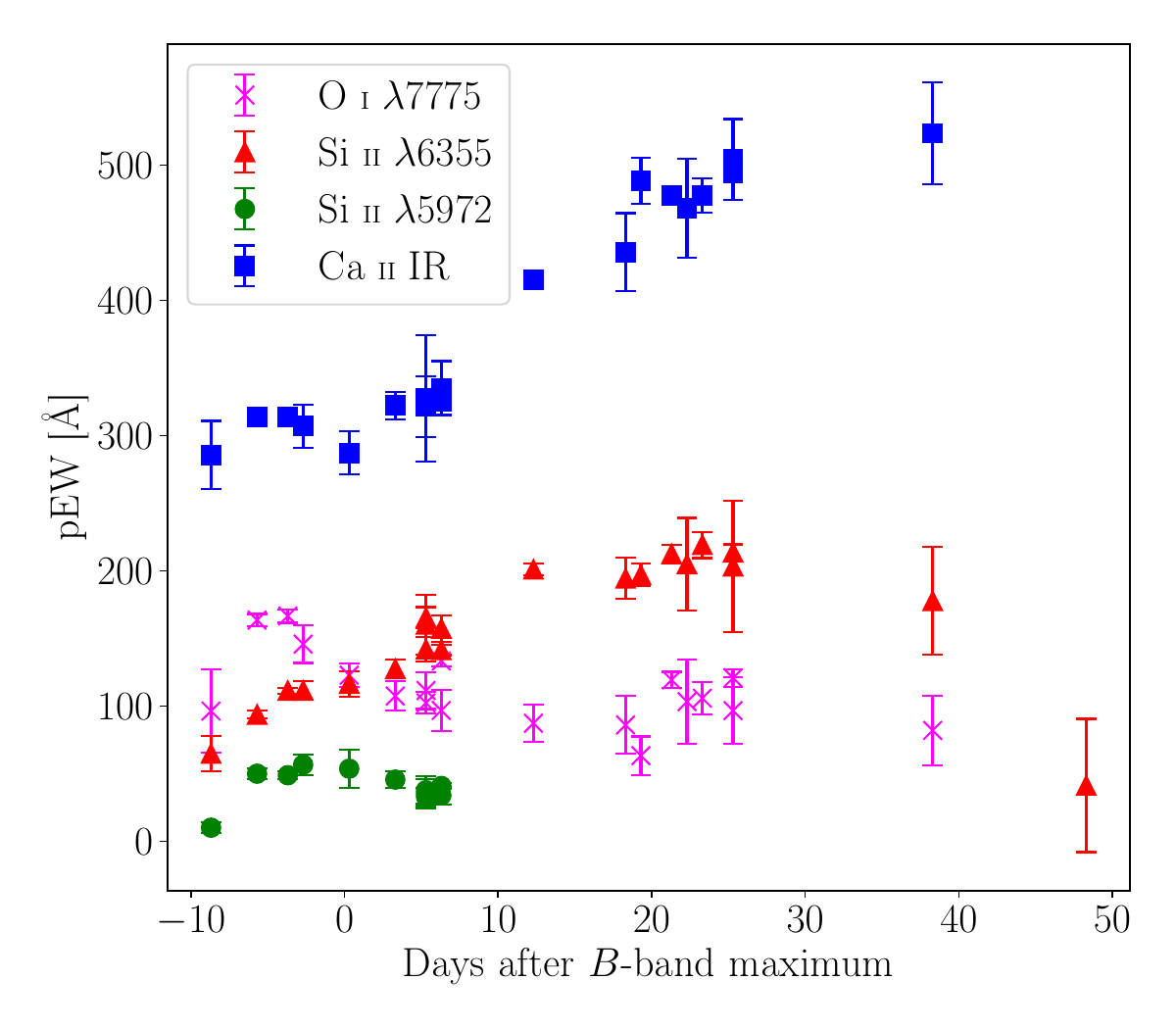}
    \caption{Evolution of the pseudo-equivalent widths.}
    \label{fig:pEW}
    \end{minipage}

\end{figure}

\subsection{Ejecta velocities and pseudo-equivalent widths}
We estimated the expansion velocities of \ion{Ca}{II}\,IR, \ion{Si}{II} $\lambda\lambda5972,6355$, \ion{O}{I} $\lambda7775$, and \ion{Ca}{II}\,H\&K using a modified version of the {\sc Spextractor} code \citep{2019PhDT.......134P}. This revised version \citep{2020ApJ...901..154B} performs spectral downsampling while conserving photon counts to reduce computational cost. The GP-based smoothing procedure was also refined to produce a more reliable representation of each spectrum, improving the measurements of expansion velocities and pEWs. Velocities are obtained from the wavelength shifts of absorption features using the relativistic Doppler formula given in Eq. (6) of \citet{2006AJ....131.1648B}, with ejecta velocities determined from the minima of the absorption lines relative to their rest wavelengths. The derived velocities are listed in Table~\ref{tab:velocities} and plotted in Fig.~\ref{fig:velocities}. 

For the  absorption lines of \ion{Si}{II} $\lambda\lambda5972,6355$, \ion{O}{I} $\lambda7773$, and \ion{Ca}{II}\,H\&K, we also measured the pseudo-equivalent widths ($pEW$). We used a simple Monte Carlo routine where the continuum points on either side of the feature were allowed to vary within $\pm 10-30$ \AA~ (depending on the spectrum) of their nominal wavelengths. For each iteration, the nearest flux values to these perturbed positions were taken to define a linear continuum, and the spectrum was normalized with respect to it. The $pEW$ was then calculated by integrating the normalized absorption profile across the chosen wavelength range. This process was repeated 1000 times, and the final $pEW$ and uncertainty were taken as the mean and standard deviation of the resulting distribution.
The $pEW$s are reported in Table~\ref{tab:velocities} (in parentheses) and whose evolution is also shown in Fig.~\ref{fig:pEW}.

\subsection{Near infrared spectra}\label{sec:NIR}
NIR spectroscopy is a powerful tool to study SN Ia physics because it allows different layers of the ejecta to be probed simultaneously \citep{2019ApJ...875L..14A}. Key spectral features, including \ion{C}{I} and \ion{Mg}{II} lines tracing unburned material and carbon burning \citep{2013ApJ...766...72H,2006ApJ...645.1392M}, and the iron-peak H-band emission, connect spectroscopic observables directly to explosion physics. The $H$-band emission is linked to $^{56}$Ni production and light-curve properties, and it also provides a probe of the temperature of the line-blanketing material \citep{1998ApJ...496..908W,2013ApJ...766...72H}.

The NIR spectra of SN~2023vjh are shown in Fig. \ref{fig:spectra_IR}. A Gaussian smoothing with $\sigma = 15-20$ has been applied.
The most prominent absorption feature is found at 8150–8300 \AA\ and corresponds to the \ion{Ca}{II} NIR triplet (8500, 8540, 8660 Å), also visible in the optical spectra shown in Fig. \ref{fig:my_spectra}. This absorption feature includes contributions from \ion{Co}{II} and \ion{Mn}{II}, however the \ion{Ca}{II} contribution is much stronger. The expansion velocities of \ion{Ca}{II} NIR are presented in Table \ref{tab:velocities}.

Since our spectra were obtained at relatively late phases, the \ion{Mg}{II} doublet (9220–9240 \AA) is no longer present. \ion{Mg}{II} forms in the outer layers of the ejecta (as a product of explosive carbon burning), and as the ejecta expand the column density decreases rapidly. As a result, the \ion{Mg}{II} feature fades and becomes undetectable a few days after maximum light (e.g. \citealt{2003ApJ...591..316M}). Then the spectrum in this region is dominated by \ion{Co}{II} (8550 \AA) and \ion{Mn}{II} (9440 \AA). However, at later times this region is dominated by blended iron-group lines, while \ion{Mn}{II} is not detectable anymore. Hence, the absorption feature seen around 9100 \AA\ is most probably due to Fe-group lines. Finally, we observe two more absorption features around 9900 \AA\ and around 10910 \AA, in all three spectra, which are due to \ion{Fe}{II} and \ion{Co}{II} respectively. The velocities corresponding to the feature minima are listed in Table~\ref{tab:NIR_vel}.
\begin{table}
    \centering
    \caption{The expansion velocities of the features associated with \ion{Fe}{II} (9900 \AA) and \ion{Co}{II} (10910 \AA) in the NIR spectra.}
    \begin{tabular}{ccc}
    \hline
        Epoch (days) & $\nu$\ion{Co}{II} ($\rm km\, s^{-1}$) & $\nu$\ion{Fe}{II} ($\rm km\, s^{-1}$)  \\
        \hline
        +16.4 & 4399 $\pm$ 540 & 2983 $\pm$ 432  \\
        +20.8 & 2401 $\pm$ 1023 & 1153 $\pm$ 374 \\
        +23.4 & 3969 $\pm$ 957 & 2295 $\pm$ 468   \\
        \hline
    \end{tabular}
    
    \label{tab:NIR_vel}
\end{table}

In Fig. \ref{fig:my_spectra} we also include NIR spectra of the 91bg-like SN~2015bo \citep{2022ApJ...928..103H} and SN~2022xkq \citep{2024ApJ...960...29P} at epochs similar to those of SN~2023vjh (a Gaussian smoothing has also been applied with $\sigma$ = 2 for SN~2015bo and 20 for SN~2022xkq). The same spectral features are present in all objects. However, \citet{2024ApJ...960...29P} report that the prominent iron-peak emission in the $H$ band, which begins to appear $\sim$5 days after $B$-band maximum, has a blue-edge velocity ($\nu_{\rm edge}$) of $-$5500 $\rm km\,s^{-1}$ at epoch +21 \citep{2019ApJ...875L..14A}. The $H$-band feature is also apparent in the spectrum of SN~2015bo. In contrast, no $H$-band break is apparent in our spectra. It is unclear whether this is due to the lower S/N at those wavelengths or the actual absence of the feature.

Finally, we examined the position of SN~2023vjh in several classification diagrams, using both its photometric and spectroscopic properties. In all cases, it lies within the locus of 91bg-like SNe Ia and cool SNe, and more specifically, it is consistent with the subclass of extreme cool SNe. A more detailed discussion of these classification diagrams is provided in Appendix~\ref{ap:class_diag}.

\begin{figure}
    \includegraphics[width=0.4\textwidth]{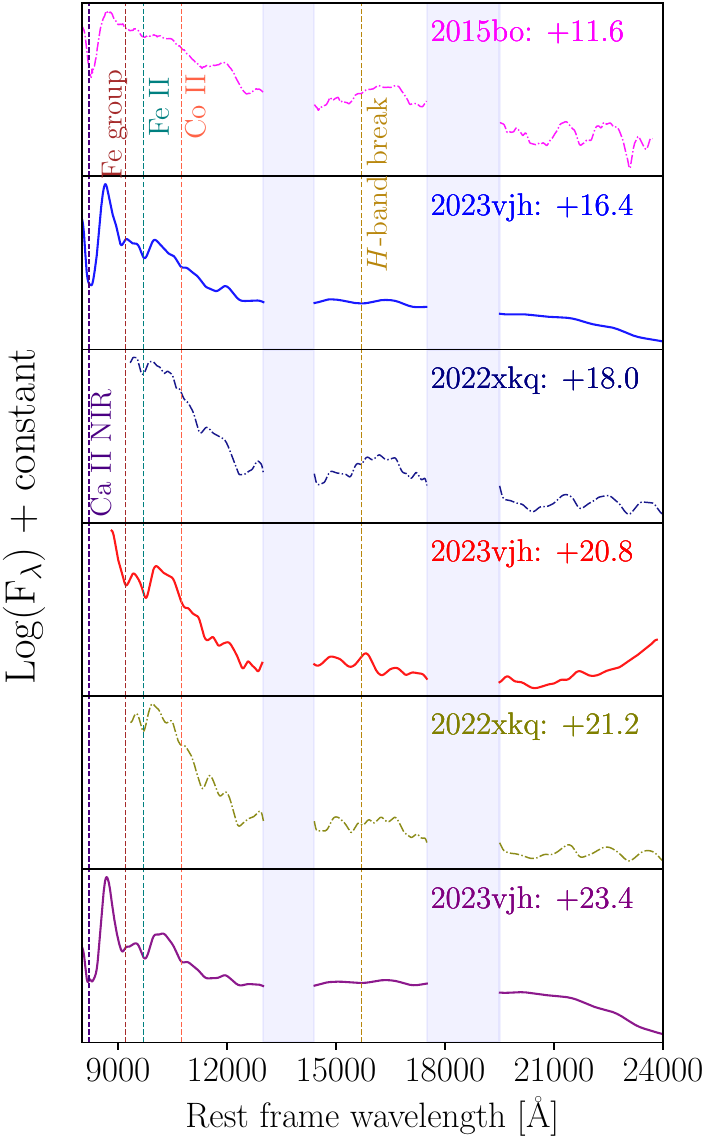}
    \caption{Near-infrared spectra of SN~2023vjh, corrected for Milky Way extinction and shifted to the rest frame, compared with those of SN~2022xkq at similar epochs. Light-blue shaded regions indicate wavelength ranges affected by strong telluric absorption, which have been excluded from the plotted spectra.}
    \label{fig:spectra_IR}
\end{figure}

\section{Summary and conclusions} \label{sum}
This paper presents observations of the 91bg-like supernova SN~2023vjh, discovered on 17 October 2023, with a rise time of 11.6 days. The best-fit light-curve parameters are $\Delta m_{15}(B)=1.89 \pm 0.01$ and $s_{BV}=0.45 \pm 0.03$, which, together with the apparent $B$-band magnitude at maximum, place SN~2023vjh firmly within the locus of 91bg-like SNe in the luminosity–width relation diagram.

Optical spectra spanning $-8.6$ to +47.3 days relative to $B$-band maximum show the typical 91bg-like features, including \ion{Si}{II}, \ion{Ca}{II}, \ion{O}{I}, and \ion{Ti}{II}. Measured expansion velocities and pseudo-equivalent widths confirm that SN~2023vjh is an extremely cool event, consistent with the definition of \citet{2013ApJ...773...53F}. In addition, late-phase NIR spectra at three epochs reveal \ion{Ca}{II} IR, \ion{Fe}{II}, and \ion{Co}{II} absorption features, complementing the optical dataset, and absence of the $H$-band break feature.

What makes SN~2023vjh particularly peculiar among 91bg-like SNe is its dimness relative to both typical events and explosion models, combined with indications of higher than usual reddening ($E(B-V){\rm _{host}} \sim$0.2–0.35). On the other hand,  the lack of Na~I~D absorption and its location 6.8 kpc from the bulge of its passive elliptical host suggest minimal interstellar extinction.

We compared the observed light curves with the Chandrasekhar-mass delayed-detonation model DDC25 and the sub-Chandrasekhar-mass central detonation model SCH2p0 from \citet{2018MNRAS.474.3931B}. We further tested the impact of circumstellar dust using the low-$R_V$, scattering-dominated extinction law of \citet{2008ApJ...686L.103G} for Milky Way- and LMC-type dust. We find that the combination of the sub-Chandrasekhar SCH2p0 model with MW-type CSM extinction provides the best overall match to the observed light curves, improving agreement in the blue bands. Residual discrepancies at red wavelengths likely reflect limitations in the model spectral energy distribution in the red/NIR regime, where strong line-forming features such as the Ca II NIR triplet can significantly affect the i-band flux.

These results show that SN~2023vjh is an unusually faint, fast-declining 91bg-like supernova. Its systematically fainter light curves, despite the lack of clear host-galaxy extinction signatures, make it a distinctive case within this subclass. A contribution from circumstellar material affecting the observed reddening cannot be excluded and may represent an interesting possibility for this event.

Finally, we used SN~2023vjh to estimate the Hubble constant following the approach of \citet{2026ApJ...998..101P}, which is tailored for fast-declining SNe~Ia. Applying both the classical Tripp relation \citep{1998A&A...331..815T} and the Color method, we find $H_0 = 71.56 \pm 3.09$ km s$^{-1}$ Mpc$^{-1}$ from the Tripp method and $H_0 = 74.60 \pm 2.22$ km s$^{-1}$ Mpc$^{-1}$ from the Color method. Both values are consistent within uncertainties with those from \citet{2026ApJ...998..101P}, with the Color method result lying slightly closer to the reference values. 


\begin{acknowledgements}
We thank the anonymous referee for the thorough review and the useful comments that helped to improve the clarity of the paper. MK acknowledges financial support from MICINN (Spain) through the programme Juan de la Cierva-Incorporación [JC2022-049447-I], from AGAUR, CSIC, MCIN and AEI 10.13039/501100011033 under projects PID2023-151307NB-I00, PIE 20215AT016, Unidad de Excelencia María de Maeztu CEX2020-001058-M, and 2021-SGR-01270
L.G. acknowledges financial support from the Spanish Ministerio de Ciencia e Innovaci\'on (MCIN) and the Agencia Estatal de Investigaci\'on (AEI) 10.13039/501100011033 under the PID2023-151307NB-I00 SNNEXT project, from Centro Superior de Investigaciones Cient\'ificas (CSIC) under projects PIE 20215AT016, ILINK23001, COOPB2304, and the program Unidad de Excelencia Mar\'ia de Maeztu CEX2020-001058-M, and from the Departament de Recerca i Universitats de la Generalitat de Catalunya through the 2021-SGR-01270 grant.
M.D. Stritzinger is funded by the Independent Research Fund Denmark (IRFD, grant number 10.46540/2032-00022B).
M.G.B. acknowledges financial support from the
Spanish Ministerio de Ciencia e Innovación (MCIN) and
the Agencia Estatal de Investigación (AEI) 10.13039/501100011033 under the
PID2023-151307NB-I00 SNNEXT project, from Centro Superior de Investigaciones Científicas (CSIC) under projects PIE 20215AT016, ILINK23001, COOPB2304, and the program Unidad de Excelencia María de Maeztu CEX2020-001058-M, and from the Departament de Recerca i Universitats de
la Generalitat de Catalunya through the 2021-SGR-01270 grant. M.G.B.’s work has been carried out within the framework of the doctoral program in Physics of the Universitat Autònoma de Barcelona.

T.-W.C. acknowledges financial support from the Yushan Fellow Program of the Ministry of Education, Taiwan (MOE-111-YSFMS-0008-001-P1), and from the National Science and Technology Council, Taiwan (NSTC 114-2112-M-008-021-MY3)
J.T.H. acknowledges support from NASA through the NASA Hubble Fellowship grant HST-HF2-51577.001-A, awarded by STScI. STScI is operated by the Association of Universities for Research in Astronomy, Incorporated, under NASA contract NAS5-26555. 

CJ acknowledges financial support from grant PRE2021-096988 funded by AEI 10.13039/501100011033 and ESF Investing in your future. CJ work has been carried out within the framework of the doctoral program in Physics of the Universitat Autònoma de Barcelona 

R.G.B. acknowledges financial support from the Severo Ochoa grant CEX2021-001131-S funded by MCIN/AEI/10.13039/501100011033 and the grant PID2022-141755NB-100.
\end{acknowledgements}
\FloatBarrier
\bibliographystyle{aa}
\bibliography{23vjh} 

\begin{appendix} 
\onecolumn
\section{Photometry log} \label{sec:logs}

\begin{table}[!htbp]
    \centering
    \small
    \caption{SWOPE photometry log}
    \begin{tabular}{llcccccc}
    \hline
         MJD   & Epoch (days)  & $B$ (mag) & $V$(mag) & $g$ (mag)& $r$ (mag)& $i$ (mag)\\
         \hline
60237.27 & $-$8.49 & 20.287 $\pm$ 0.043 & 19.450 $\pm$ 0.026 & 19.798 $\pm$ 0.024 & 19.134 $\pm$ 0.019 & 19.270 $\pm$ 0.030 \\
60238.29 & $-$7.47 & 19.962 $\pm$ 0.032 & 19.116 $\pm$ 0.021 & 19.489 $\pm$ 0.019 & 18.789 $\pm$ 0.016 & 18.870 $\pm$ 0.020 \\
60239.31 & $-$6.45 & 19.687 $\pm$ 0.025 & 18.818 $\pm$ 0.017 & 19.227 $\pm$ 0.017 & 18.528 $\pm$ 0.017 & 18.590 $\pm$ 0.019 \\
60240.35 & $-$5.41 & 19.543 $\pm$ 0.024 & 18.605 $\pm$ 0.017 & 19.043 $\pm$ 0.019 & 18.282 $\pm$ 0.013 & 18.357 $\pm$ 0.020 \\
60241.3 & $-$4.46 & 19.397 $\pm$ 0.045 & 18.441 $\pm$ 0.022 & 18.864 $\pm$ 0.019 & 18.094 $\pm$ 0.015 & 18.220 $\pm$ 0.018 \\
60242.31 & $-$3.45 & 19.208 $\pm$ 0.030 & 18.254 $\pm$ 0.019 & 18.700 $\pm$ 0.022 & 17.970 $\pm$ 0.015 & 18.026 $\pm$ 0.017 \\
60243.27 & $-$2.49 & 19.146 $\pm$ 0.027 & 18.132 $\pm$ 0.012 & 18.614 $\pm$ 0.017 & 17.819 $\pm$ 0.016 & 17.918 $\pm$ 0.017 \\
60244.3 & $-$1.46 & 18.956 $\pm$ 0.034 & 18.068 $\pm$ 0.035 & 18.571 $\pm$ 0.024 & 17.710 $\pm$ 0.018 & 17.770 $\pm$ 0.017 \\
60245.32 & $-$0.44 & 18.783 $\pm$ 0.031 & 18.036 $\pm$ 0.016 & 18.571 $\pm$ 0.027 & 17.634 $\pm$ 0.022 & 17.710 $\pm$ 0.023 \\
60250.33 & 4.57 & 19.255 $\pm$ 0.037 & 18.073 $\pm$ 0.029 & 18.709 $\pm$ 0.059 & 17.613 $\pm$ 0.019 & 17.625 $\pm$ 0.022 \\
60251.33 & 5.57 & 19.458 $\pm$ 0.088 & 18.188 $\pm$ 0.028 & 18.921 $\pm$ 0.042 & 17.641 $\pm$ 0.016 & 17.661 $\pm$ 0.014 \\
60252.22 & 6.46 & 19.598 $\pm$ 0.025 & 18.267 $\pm$ 0.011 & 19.075 $\pm$ 0.015 & 17.717 $\pm$ 0.013 & 17.713 $\pm$ 0.013 \\
60253.32 & 7.56 & 19.958 $\pm$ 0.077 & 18.399 $\pm$ 0.020 & 19.253 $\pm$ 0.033 & 17.835 $\pm$ 0.012 & 17.754 $\pm$ 0.015 \\
60254.3 & 8.54 & 19.978 $\pm$ 0.052 & 18.510 $\pm$ 0.018 & 19.317 $\pm$ 0.022 & 17.898 $\pm$ 0.010 & 17.821 $\pm$ 0.011 \\
60255.27 & 9.51 & 20.120 $\pm$ 0.033 & 18.641 $\pm$ 0.015 & 19.529 $\pm$ 0.021 & 18.007 $\pm$ 0.011 & 17.879 $\pm$ 0.012 \\
60256.23 & 10.47 & 20.142 $\pm$ 0.036 & 18.766 $\pm$ 0.015 & 19.624 $\pm$ 0.022 & 18.088 $\pm$ 0.013 & 17.926 $\pm$ 0.011 \\
60260.28 & 14.52 & 20.693 $\pm$ 0.056 & 19.157 $\pm$ 0.024 & 20.128 $\pm$ 0.033 & 18.489 $\pm$ 0.015 & 18.242 $\pm$ 0.016 \\
60261.19 & 15.43 & 20.755 $\pm$ 0.047 & 19.260 $\pm$ 0.021 & 20.093 $\pm$ 0.031 & 18.580 $\pm$ 0.015 & 18.418 $\pm$ 0.031 \\
60262.24 & 16.48 & 20.767 $\pm$ 0.044 & 19.345 $\pm$ 0.023 & 20.204 $\pm$ 0.031 & 18.649 $\pm$ 0.017 & 18.418 $\pm$ 0.017 \\
60263.32 & 17.56 &  $\cdots$ & 19.390 $\pm$ 0.094 &  $\cdots$ &  $\cdots$ &  $\cdots$ \\
60264.19 & 18.43 &  $\cdots$ & 19.492 $\pm$ 0.047 & 20.313 $\pm$ 0.049 & 18.848 $\pm$ 0.020 & 18.623 $\pm$ 0.021 \\
60265.18 & 19.42 & 20.991 $\pm$ 0.125 & 19.551 $\pm$ 0.048 & 20.344 $\pm$ 0.055 & 18.902 $\pm$ 0.022 & 18.670 $\pm$ 0.031 \\
60266.17 & 20.41 & 21.021 $\pm$ 0.091 & 19.639 $\pm$ 0.033 & 20.358 $\pm$ 0.056 & 18.974 $\pm$ 0.021 & 18.803 $\pm$ 0.030 \\
60267.21 & 21.45 & 21.149 $\pm$ 0.065 & 19.739 $\pm$ 0.027 & 20.465 $\pm$ 0.040 & 19.062 $\pm$ 0.022 & 18.869 $\pm$ 0.025 \\
60268.18 & 22.42 & 20.908 $\pm$ 0.051 & 19.725 $\pm$ 0.031 &  $\cdots$ & 19.102 $\pm$ 0.023 & 18.950 $\pm$ 0.026 \\
60269.22 & 23.46 & 21.076 $\pm$ 0.056 & 19.826 $\pm$ 0.028 & 20.659 $\pm$ 0.047 & 19.202 $\pm$ 0.027 & 19.014 $\pm$ 0.027 \\
60280.21 & 34.45 & 21.414 $\pm$ 0.162 & 20.249 $\pm$ 0.037 & 20.964 $\pm$ 0.060 & 19.713 $\pm$ 0.026 & 19.623 $\pm$ 0.024 \\
60281.18 & 35.42 & 21.484 $\pm$ 0.105 & 20.330 $\pm$ 0.040 & 20.962 $\pm$ 0.042 & 19.751 $\pm$ 0.022 & 19.620 $\pm$ 0.025 \\
60283.18 & 37.42 & 21.457 $\pm$ 0.062 & 20.323 $\pm$ 0.040 & 20.928 $\pm$ 0.041 & 19.839 $\pm$ 0.027 & 19.703 $\pm$ 0.026 \\
60285.18 & 39.42 & 21.549 $\pm$ 0.064 & 20.396 $\pm$ 0.047 & 21.042 $\pm$ 0.058 & 19.925 $\pm$ 0.034 & 19.812 $\pm$ 0.030 \\
60287.17 & 41.41 & 21.513 $\pm$ 0.071 & 20.454 $\pm$ 0.039 & 21.065 $\pm$ 0.053 & 20.021 $\pm$ 0.032 & 19.925 $\pm$ 0.034 \\
60289.15 & 43.39 & 21.756 $\pm$ 0.088 & 20.539 $\pm$ 0.054 & 21.152 $\pm$ 0.060 & 20.118 $\pm$ 0.032 & 20.001 $\pm$ 0.036 \\
60291.13 & 45.37 & 21.738 $\pm$ 0.083 &  $\cdots$ & 21.236 $\pm$ 0.067 & 20.165 $\pm$ 0.037 & 20.036 $\pm$ 0.037 \\
\hline
    \end{tabular}
    \label{tab:photometry_poise}
\end{table}

\begin{longtable}{llccccc}
\caption{ZTF $r,\,g$ and ATLAS $c,\,o$, photometry}
\label{tab:photometry_atlas_ztf}\\

\hline
MJD & Epoch (days) & ZTF $r$ (mag) & ZTF $g$ (mag)& ATLAS $o$ (mag) & ATLAS $c$ (mag)\\
\hline
\endfirsthead

\hline
MJD & Epoch (days) & ZTF $r$ (mag) & ZTF $g$ (mag) & ATLAS $o$ (mag)& ATLAS $c$ (mag)\\
\hline
\endhead
60234.38 &$ -$11.38 & 20.639 $\pm$ 0.152 & 21.105 $\pm$ 0.203 & $\cdots$ & $\cdots$ \\
60235.49 & $-$10.27 &  $\cdots$ &  $\cdots$ & $\cdots$& 19.720 $\pm$ 0.139 \\
60236.38 & $-$9.38 & 19.609 $\pm$ 0.063 & 20.180 $\pm$ 0.089 & $\cdots$ & $\cdots$\\
60237.47 & $-$8.29 & $\cdots$ &  $\cdots$ &  $\cdots$ & 19.510 $\pm$ 0.155 \\
60238.39 & $-$7.37 & 18.707 $\pm$ 0.029 & 19.387 $\pm$ 0.044 & $\cdots$ & $\cdots$ \\
60239.47 & $-$6.29 & $\cdots$ & $\cdots$ & $\cdots$& 18.779 $\pm$ 0.057 \\
60241.41 & $-$4.35 & 18.107 $\pm$ 0.024 & 18.836 $\pm$ 0.046 & 18.176 $\pm$ 0.049 & $\cdots$ \\
60242.2 &$ -$3.56 & $\cdots$ & $\cdots$ & 18.080 $\pm$ 0.074 & $\cdots$ \\
60244.35 &$ -$1.41 & 17.660 $\pm$ 0.022 & 18.471 $\pm$ 0.057 & $\cdots$& $\cdots$ \\
60245.37 & $-$0.39 & $\cdots$ & $\cdots$ & 17.968 $\pm$ 0.313 & $\cdots$\\
60246.36 & 0.6 & 17.614 $\pm$ 0.039 & 18.420 $\pm$ 0.085 & $\cdots$ & $\cdots$ \\
60249.51 & 3.75 & $\cdots$ & 18.727 $\pm$ 0.158 & $\cdots$ & $\cdots$ \\
60251.41 & 5.65 & 17.624 $\pm$ 0.016 & 18.990 $\pm$ 0.056 & 17.635 $\pm$ 0.029 &  $\cdots$\\
60253.34 & 7.58 & 17.750 $\pm$ 0.016 & 19.193 $\pm$ 0.046 & 17.826 $\pm$ 0.036 & $\cdots$ \\
60254.21 & 8.45 & $\cdots$ & $\cdots$ & 17.834 $\pm$ 0.053 & $\cdots$ \\
60257.36 & 11.6 & 18.328 $\pm$ 0.055 & 19.928 $\pm$ 0.099 & $\cdots$ & 19.014 $\pm$ 0.076 \\
60259.42 & 13.66 & 18.292 $\pm$ 0.025 & 20.141 $\pm$ 0.127 & $\cdots$ & 19.119 $\pm$ 0.080 \\
60260.29 & 14.53 & $\cdots$ & $\cdots$ & 18.353 $\pm$ 0.129 & $\cdots$ \\
60261.39 & 15.63 & 18.493 $\pm$ 0.025 & 20.128 $\pm$ 0.112 & $\cdots$ & 19.530 $\pm$ 0.117 \\
60263.45 & 17.69 & 18.772 $\pm$ 0.035 & $\cdots$& $\cdots$ & 19.543 $\pm$ 0.148 \\
60265.49 & 19.73 &  $\cdots$ & $\cdots$ & $\cdots$ & 19.697 $\pm$ 0.140 \\
60266.19 & 20.43 &  $\cdots$ & $\cdots$ & 18.999 $\pm$ 0.115 & $\cdots$ \\
60267.49 & 21.73 &  $\cdots$ & $\cdots$ & $\cdots$ & 19.979 $\pm$ 0.227 \\
60269.51 & 23.75 &  $\cdots$ & $\cdots$ & $\cdots$ & 19.835 $\pm$ 0.173 \\
60281.44 & 35.68 &  $\cdots$ & $\cdots$ & 19.617 $\pm$ 0.187 & $\cdots$ \\
60282.42 & 36.66 & $\cdots$ & $\cdots$ & 19.675 $\pm$ 0.244 & $\cdots$ \\
60283.41 & 37.65 &  $\cdots$ & $\cdots$ & $\cdots$ & 20.703 $\pm$ 0.330 \\
60284.22 & 38.46 & 20.067 $\pm$ 0.111 & 21.329 $\pm$ 0.315 & 19.812 $\pm$ 0.116 & $\cdots$ \\
60285.39 & 39.63 & $\cdots$ & $\cdots$ & 19.842 $\pm$ 0.159 & 20.322 $\pm$ 0.232 \\
60287.37 & 41.61 & $\cdots$ & $\cdots$& 19.544 $\pm$ 0.188 & 20.575 $\pm$ 0.222 \\
60288.33 & 42.57 & 20.276 $\pm$ 0.171 & 21.000 $\pm$ 0.296 & 20.096 $\pm$ 0.217 &  $\cdots$ \\
60290.27 & 44.51 & 20.273 $\pm$ 0.125 & 20.871 $\pm$ 0.178 & $\cdots$ &  $\cdots$ \\
60291.42 & 45.66 & $\cdots$ & $\cdots$ & 20.104 $\pm$ 0.206 & $\cdots$ \\
60292.3 & 46.54 & 20.402 $\pm$ 0.097 & 23.062 $\pm$ 0.158 & 19.785 $\pm$ 0.172 & $\cdots$ \\
60293.23 & 47.47 & 20.569 $\pm$ 0.170 & 22.971 $\pm$ 0.296 &  $\cdots$& 20.461 $\pm$ 0.252 \\
60294.22 & 48.46 & 20.529 $\pm$ 0.118 & 22.910 $\pm$ 0.242 & $\cdots$ & $\cdots$\\
60295.41 & 49.65 & $\cdots$ & $\cdots$& 20.032 $\pm$ 0.201 & $\cdots$ \\
60308.28 & 62.52 & 21.533 $\pm$ 0.358 & $\cdots$& 20.272 $\pm$ 0.338 & $\cdots$\\
60312.36 & 66.6 & $\cdots$ & $\cdots$ & 20.427 $\pm$ 0.261 & $\cdots$\\
\hline

\end{longtable}
\twocolumn

\begin{figure}[t]
    \centering
    \includegraphics[width=0.4\textwidth]{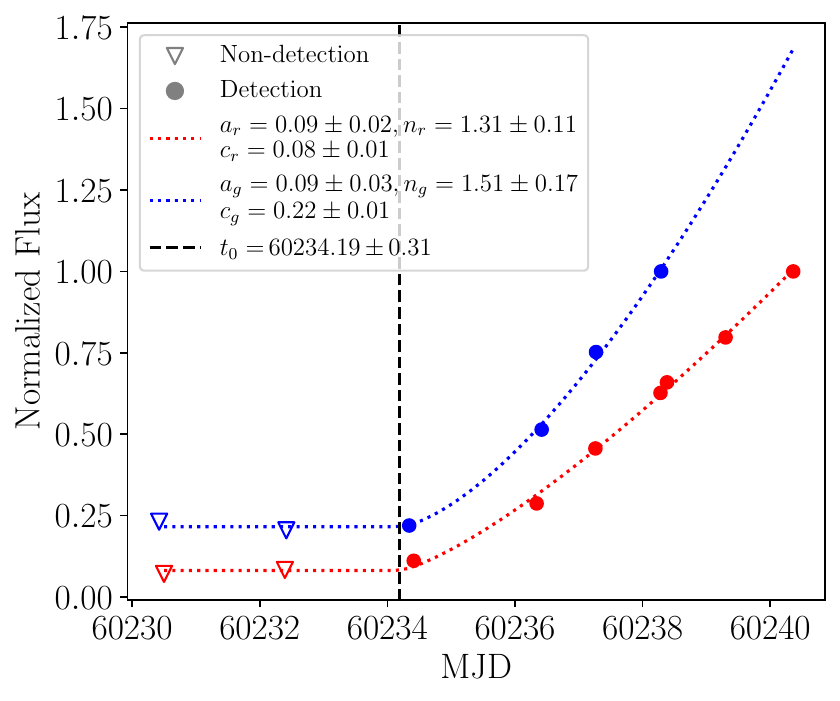}
    \caption{Power-law fitting of the early rise of the light curves $g$, $r$, $ZTF-g$, $ZTF-r$ of SN~2023vjh. We fit simultaneously the function of eq. \ref{eq:exp_date_eq} for the red and green bands, and we define the explosion date ($t_0$).}
    \label{fig:exp_time}
\end{figure}
\begin{figure}    
\makebox[\columnwidth][c]{
        \includegraphics[width=0.7\columnwidth]{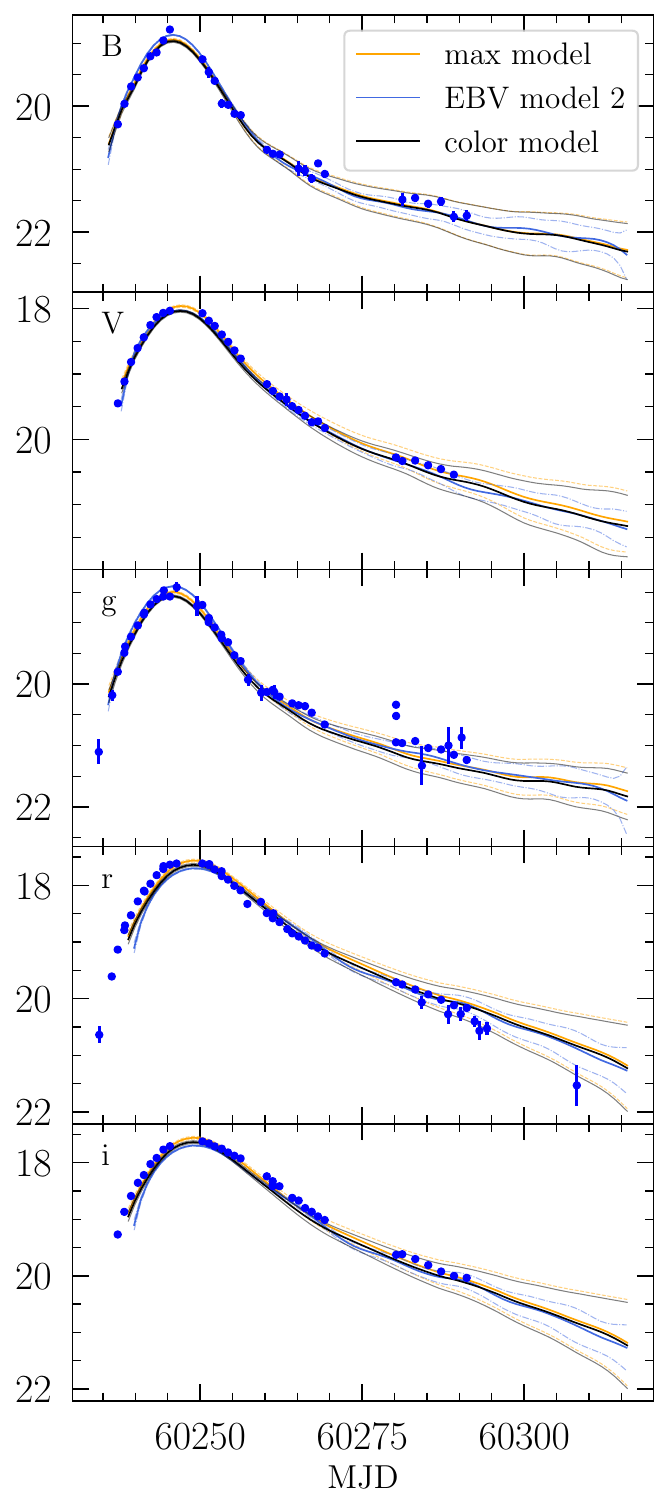}
}
        \caption{Multi-band light curves of \texttt{SNooPy} fit of SN~2023vjh using the {\texttt{max\_model}} (dashed, red line), {\texttt{EBV\_model2}} (dashed-dotted, green line), and {\texttt{color\_model}} (dashed, black line). All the fits have been done with a 1991bg-like template.        
        }
        \label{fig:snpy_fit_maxmodel}

    \vspace{0.4em}
\end{figure}

\begin{figure}
     \centering
     \small
    \includegraphics[width=0.4\textwidth]{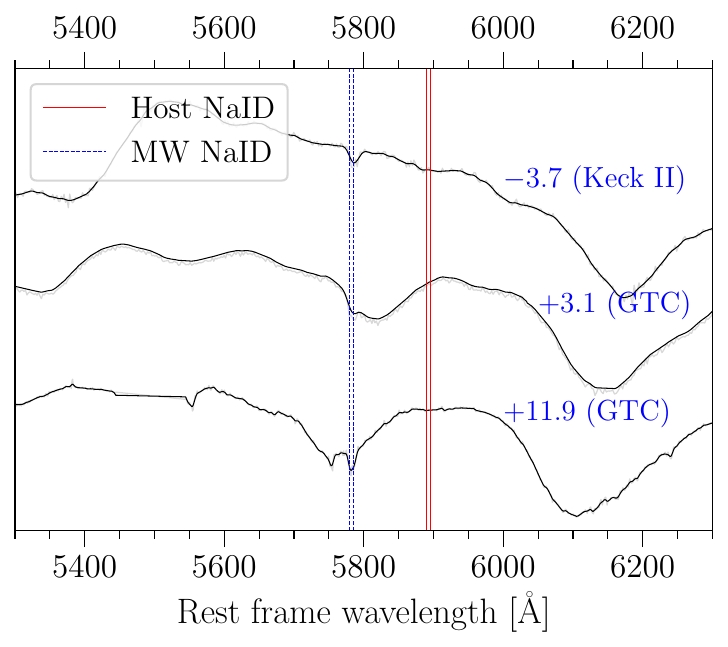}
     \caption{Spectra of SN~2023vjh where the position of Na~I~D absorption lines for Milky Way and the host galaxy are indicated.}
     \label{fig:NaID}
 \end{figure}

\section{Spectra log and spectroscopic properties}
\setlength{\tabcolsep}{3pt}
\begin{table}[!h]
    \caption{Spectroscopic log}

    \small
    \begin{tabular}{lllll}
    \hline
         UT date    & MJD   & Epoch  & Telescope & Range/Grism \\
         \hline
         2023-10-20 & 60237  & $-$8.6  & UH88      & 3000 - 10000\\
         2023-10-23 & 60240  & $-$5.8  & GTC       & R1000B/R1000R\\
         2023-10-25 & 60242  & $-$3.7  & Keck II     & 3000 - 10000\\
         2023-10-26 & 60243  & $-$2.7  & UH88      & 3000 - 10000\\
         2023-10-29 & 60246  & +0.2  & UH88      & 3000 - 10000\\
         2023-11-01 & 60252  & +3.1  & GTC       & R1000B/R1000R\\
         2023-11-01 & 60249  & +3.1  & NOT 2.6m  & gr4\\
         2023-11-03 & 60251  & +5.1  & UH88      & 3000 - 10000\\
         2023-11-03 & 60251  & +5.1  & LT        & blue\\
         2023-11-03 & 60251  & +5.1  & ESO NTT   & gr11+gr16\\
         2023-11-04 & 60252  & +6.1  & UH88      & 3000 - 10000\\
         2023-11-10 & 60258  & +11.9 & GTC       & R1000B/R1000R\\
         2023-11-16 & 60264  & +17.9 & ESO NTT   & gr13 \\
         2023-11-17 & 60265  & +18.9 & ESO NTT   & gr16 \\
         2023-11-19 & 60267  & +20.8 & Keck II     & 3000 - 10000\\
         2023-11-20 & 60268  & +21.8 & UH88      & 3000 - 10000\\
         2023-11-21 & 60269  & +22.8 & Baade     & 4230 - 9418\\
         2023-11-23 & 60271  & +24.8 & UH88       & 3000 - 10000\\
         2023-11-23 & 60271  & +24.7 & GTC      & R1000B/R1000R\\
         2023-12-06 & 60284  & +37.5 & UH88      & 3000 - 10000\\
         2023-12-15 & 60294  & +47.3 & CAHA Tel. 3.5m  & V500 - IFU\\
         \hline
    \end{tabular}
    \label{tab:spectra}
\end{table}

\begin{table*}
    \centering
    \tiny
    \begin{tabular}{llcccccc}
    \hline
    
MJD   & Epoch  & $v$\ion{Ca}{II} IR ($km\, s^{-1}$) & $v$\ion{Si}{II} $\lambda$5972  ($km\, s^{-1}$)& $v$\ion{Si}{II} $\lambda$6355 ($km\, s^{-1}$) & $v$\ion{O}{I} $\lambda$7775 ($km\, s^{-1}$) & $v$\ion{Ca}{II} H\&K ($km\, s^{-1}$) & Telescope \\  
    &    & ($pEW$ [\AA]) & ($pEW$ [\AA])  & ($pEW$ [\AA]) & ($pEW$ [\AA])  & ($pEW$ [\AA]) \\   
\hline
60237 & $-$8.66 & 16114 $\pm$ 590 & 13836 $\pm$ 745 &  14903 $\pm$ 334 & 15035 $\pm$ 382 & $-$ & UH88\\
      &       & (285.7 $\pm$ 25.3)  & (10.2 $\pm$ 3.9) & (65 $\pm$ 13)  & (96.4 $\pm$ 30.8) & $-$ & \\

60240 & $-$5.72 & 14825 $\pm$ 359    & 11847 $\pm$ 181 &  12639 $\pm$ 264 & 11868 $\pm$ 265 & 16048 $\pm$ 341 & GTC\\
      &       & (313.6 $\pm$ 5.2)  &  (50.1  $\pm$ 4.1) & (93.9 $\pm$ 3.1)  & (163.7  $\pm$ 4.5) &  (91.5  $\pm$ 20.4) &\\

60242 & $-$3.76 & 14550 $\pm$ 441 & 10760 $\pm$ 288 &  12178 $\pm$ 237 & 11939 $\pm$ 102 & 15592 $\pm$ 497 & Keck\\
      &       & (313.9 $\pm$ 6.8) & (49.0 $\pm$ 2.8) &  (111.5 $\pm$ 2.2) &  (166.7  $\pm$ 5.0) & (108.6 $\pm$ 5.7) &\\

60243 & $-$2.78 & 13431 $\pm$ 970  & 10475 $\pm$ 295 &  11208 $\pm$ 284 & 13404 $\pm$ 407 & 13969 $\pm$ 2494 & UH88\\
      &       & (307.3 $\pm$ 16.0) & (56.8 $\pm$ 7.6) & (111.6 $\pm$ 7.1) &  (146.0 $\pm$ 14.0) &  (307.3 $\pm$ 16.0) & \\

60246 & 0.17 & 12141 $\pm$ 1074 & 9067  $\pm$ 425 &  10269 $\pm$ 498  & 13244 $\pm$ 414 & 24487 $\pm$ 8060 & UH88\\
      &      & (287.3 $\pm$ 15.8) &  (53.8 $\pm$ 14.1) & (116.7 $\pm$ 9.4)  &  (122.9  $\pm$ 8.8) & (287.3 $\pm$ 15.8)& \\

60249 & 3.11 & 13601 $\pm$ 830  & 8346  $\pm$ 187 & 10717 $\pm$ 251 & 13036 $\pm$ 427 & 16952 $\pm$ 772 & NOT \\
      &      & (322.3 $\pm$ 10.3) & (45.7  $\pm$ 6.2) & (127.5 $\pm$ 6.9) &  (107.5 $\pm$ 10.9)  & (216.9 $\pm$ 246.6)&\\
60249 & 3.11 & 13732  $\pm$ 1135 & 7894 $\pm$ 125 & 9871  $\pm$ 263 & 11448 $\pm$ 99  & 13803 $\pm$ 468 & GTC\\
     &     &  (325.6 $\pm$ 4.3) & (41.0 $\pm$ 2.3) & (141.6 $\pm$ 3.8) & (133.4 $\pm$ 4.0) & (88.4 $\pm$ 14.4)& \\

60251 & 5.07 & 11534 $\pm$ 547  & 7350 $\pm$ 410 & 9999 $\pm$ 416 & 11019 $\pm$ 178 & 32300 $\pm$ 6373 & UH88\\
      &      & (321.5 $\pm$ 22.7) & (33.1 $\pm$ 6.8) &  (142.1 $\pm$ 9.1)  & (102.48  $\pm$ 8.0) &  (235.2 $\pm$  7.7) & \\

60251 & 5.07 & $-$ &    $-$ &     8831 $\pm$ 889 & 11767 $\pm$ 760 & $-$ & LT\\
      &      & $-$ &    $-$ &     (160.5 $\pm$ 22.1) & $-$ & $-$ &\\

60251 & 5.07 & 11459 $\pm$ 865 & 6767 $\pm$ 385 &  9736 $\pm$ 274 & 11558 $\pm$ 225 & 44450 $\pm$ 974 & ESO NTT\\
      &      &  (327.6 $\pm$ 46.8)  &  (35.4 $\pm$ 10.7)  &  (165.6  $\pm$ 7.7) &  (111.6 $\pm$ 13.7) &  (268.9 $\pm$ 88.9) & \\

60252 & 6.05 & 13155 $\pm$ 844 & 7208 $\pm$ 369 & 9864 $\pm$ 332 & 11019 $\pm$ 272 & $-$ & UH88\\
      &      &  (335.3 $\pm$  19.9)   & (34.1 $\pm$ 6.6)  &  (157.4 $\pm$  10.0)   &  (96.9 $\pm$ 15.1) & $-$& \\

60258 & 11.94 & 12773 $\pm$ 352 & $-$ & 8735 $\pm$ 146 & 11177 $\pm$ 75 & 10628 $\pm$ 583 & GTC \\
     &        & (415.5 $\pm$ 2.6) & $-$ & (201.2 $\pm$ 4.2) & (87.5 $\pm$ 13.7) &   (73.0 $\pm$ 53.9)& \\

60264 & 17.83 & 12083 $\pm$ 615 & $-$ &  8049 $\pm$ 562 & 10267 $\pm$ 708 & $-$ & ESO NTT \\
      &   & (435.8 $\pm$ 28.8) & $-$ &  (194.4 $\pm$ 15.2) & (86.2 $\pm$ 21.4) & $-$ &\\

60265 & 18.81 & 9926 $\pm$ 1129   & $-$ &  7976 $\pm$ 448 & 10259 $\pm$ 356 & 11769 $\pm$ 536 & ESO NTT \\
          &   & (488.4 $\pm$ 17.2)  & $-$ & (197.1 $\pm$ 8.4)  &  (63.4 $\pm$ 14.2) &  (99.4 $\pm$ 34.2)&\\

60267 & 20.77 & 12216 $\pm$ 276 & $-$ &  7383 $\pm$ 351 & 11730 $\pm$ 81 & 12316 $\pm$ 313 & Keck\\
      &       & (477.7 $\pm$ 6.7) & $-$ &  (212.6 $\pm$ 6.8) & (119.4 $\pm$ 6.1) &  (94.1 $\pm$ 7.8) & \\

60268 & 21.75 & 14173 $\pm$ 938 & $-$ &  7306 $\pm$ 419 & 9801 $\pm$ 268 & $-$ & UH88 \\
       &      &  (468.4 $\pm$ 36.6) & $-$ & (205.1  $\pm$ 34.1) &  (103.4 $\pm$ 31.4) &  $-$ & \\

60269 & 22.73 & 12709 $\pm$ 780  & $-$ &  7368 $\pm$ 198 & 10726 $\pm$ 124 & $-$ & Baade \\
      &       & (477.4 $\pm$ 12.7) & $-$ &  (219.3 $\pm$ 9.7) & (105.9 $\pm$ 11.8) & $-$ &\\

60271 & 24.7 & 12705 $\pm$ 1250 & $-$ &  7649 $\pm$ 916 & 9644 $\pm$ 529 & 9209 $\pm$ 4453 & UH88 \\
      &      & (504.2 $\pm$ 29.9) & $-$ &  (203.4 $\pm$ 48.7)  &  (96.7 $\pm$ 24.5) &      $-$  &    \\

60271 & 24.7 & 10959 $\pm$ 757 &  $-$ &  7312 $\pm$ 136 & 11152 $\pm$ 116 & 11983 $\pm$ 269 & GTC \\
       &     & (494.1 $\pm$  6.8)  &  $-$ &  (213.7   $\pm$ 6.0)    &  (121.1 $\pm$ 6.4) & (94.9  $\pm$ 16.9) & \\

60284 & 37.45 & 10499 $\pm$ 732 & $-$ &  5298 $\pm$ 1368 & 7872 $\pm$ 1828 & $-$ & UH88\\
      &       & (523.7 $\pm$ 37.4)  & $-$ & (177.9 $\pm$ 39.8)  & (82.1  $\pm$ 25.7) & $-$ &\\

60294 & 47.26 & $-$ &   $-$ &  1134 $\pm$ 2430 & 33899 $\pm$ 1480 & 11487 $\pm$ 771  & CAHA\\
      &       & $-$ &   $-$ &  (41.4 $\pm$ 49.3) &  $-$  & (113.6 $\pm$ 536.2) &\\

\hline
\end{tabular}
\caption{Expansion velocities and pseudo-equivalent widths}
    \label{tab:velocities}
\end{table*}

\section{Position of SN~2023vjh in classification diagrams} \label{ap:class_diag}
In previous surveys, the spectroscopic properties of SNe~Ia near maximum light have been used to classify them into different subtypes and to identify peculiar or uncommon events (e.g., \citealt{2006PASP..118..560B,2012AJ....143..126B,2013ApJ...773...53F,2024ApJ...967...20M}). Here, we reproduce some of these classification diagrams in order to examine the position of SN~2023vjh within this context. Based on these earlier works, 91bg-like SNe Ia are expected to fall within the region occupied by “cool” SNe Ia, which are characterized by lower photospheric temperatures and spectra dominated by strong low-ionization features such as Fe II, Ti II, and Ca II \citep{1995ApJ...455L.147N,1997MNRAS.284..151M,2006PASP..118..560B}. 

In Fig.~\ref{fig:CL_vel}, we show the temporal evolution of the \ion{Si}{II} $\lambda6355$ expansion velocity for SN~2023vjh (black line), alongside the sample of cool SNe from \citet{2013ApJ...773...53F}. The \ion{Si}{II} $\lambda6355$ velocity in SN~2023vjh follows the same decline trend as the cool SNe~Ia (shown in brown), which typically exhibit faster velocity decreases than core-normal SNe~Ia, particularly after $B$-band maximum. For comparison, we also include the mean \ion{Si}{II} $\lambda6355$ velocity evolution and its uncertainties for core-normal SNe from \citet{2024ApJ...967...20M}, highlighting the clear difference between the two populations.
 \begin{figure*}[htbp]
    \centering

    \begin{subfigure}{\columnwidth}
        \centering
        \includegraphics[width=\linewidth]{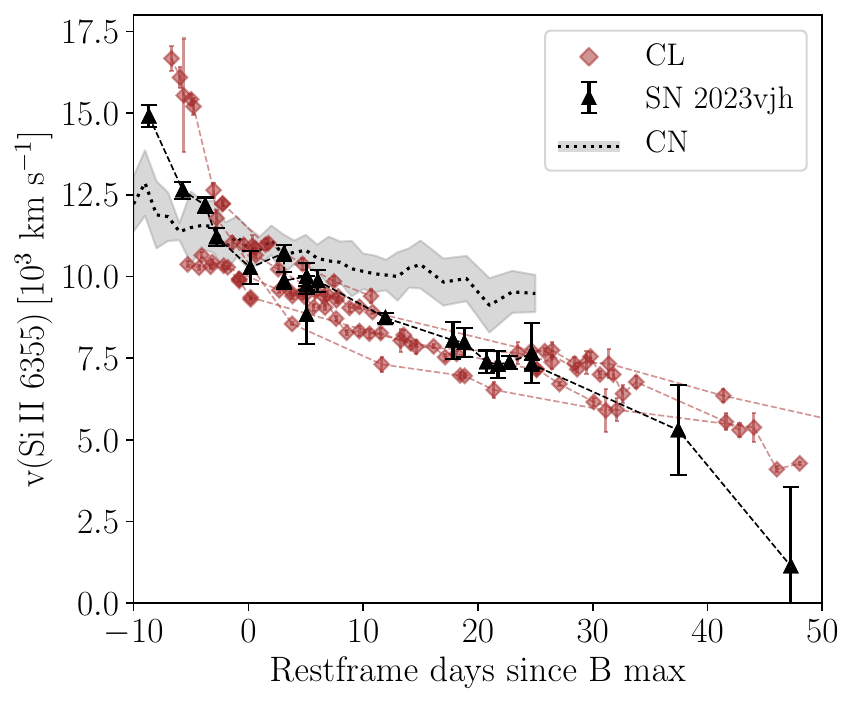}
        \caption{\ion{Si}{II} $\lambda6355$ expansion velocity of SN~2023vjh (black triangles) and other cool (CL) SNe. The black-dotted line indicates the mean \ion{Si}{II} $\lambda6355$ velocity of core normal (CN) SNe and the gray zone its uncertainty \citep{2024ApJ...967...20M}.}
        \label{fig:CL_vel}
    \end{subfigure}
    \hfill
    \begin{subfigure}{\columnwidth}
        \centering
        \includegraphics[width=\linewidth]{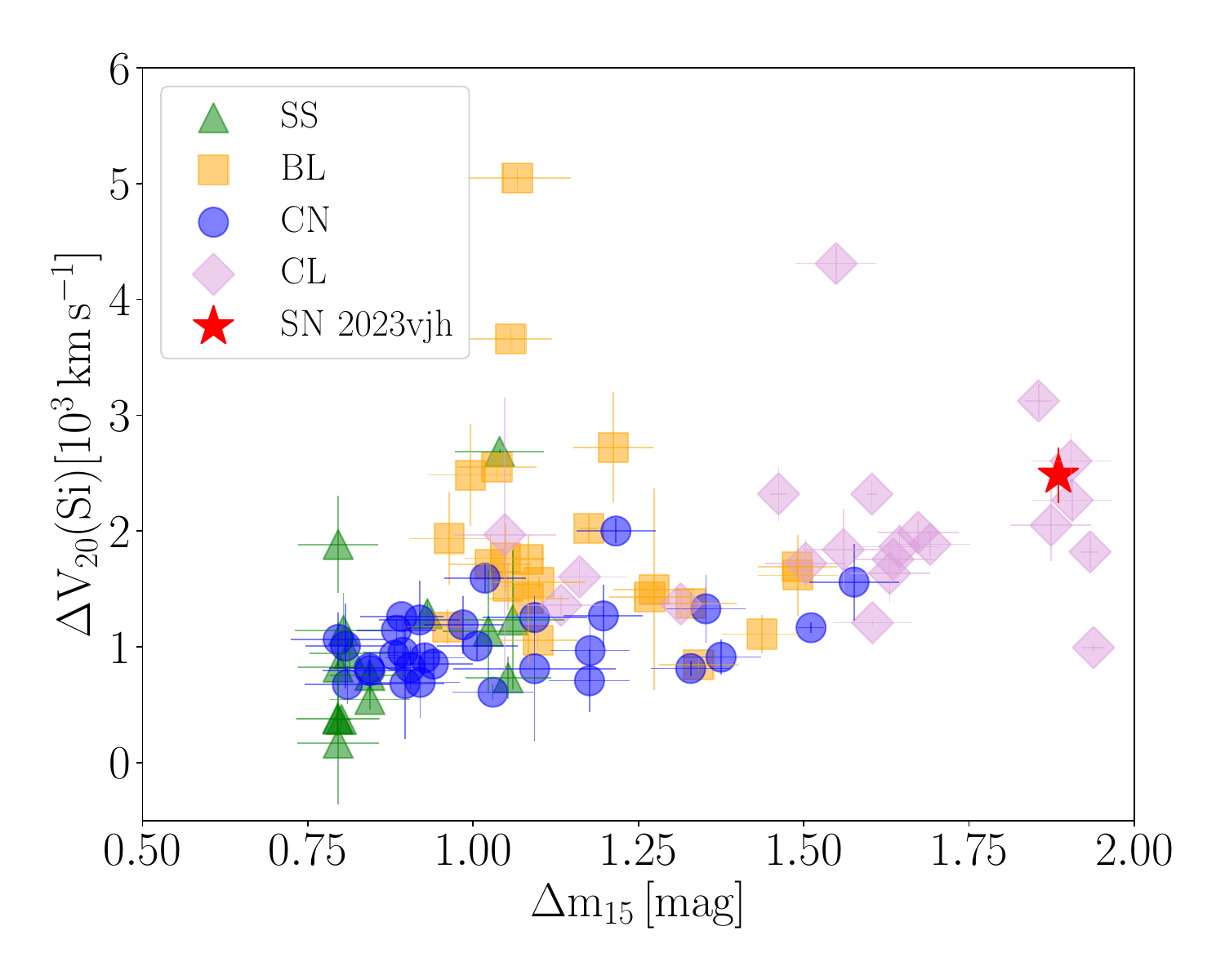}
        \caption{The \ion{Si}{II} $\lambda6355$ velocity decline rate as a function of the $\rm \Delta m_{15}$. The red star indicates the position of SN~2023vjh, while the rest of the sources have been obtained by \citet{2024ApJ...967...20M}.}
        \label{fig:dv20}
    \end{subfigure}

\end{figure*}

\begin{figure*}[htbp]
    \centering
    \includegraphics[width=\textwidth]{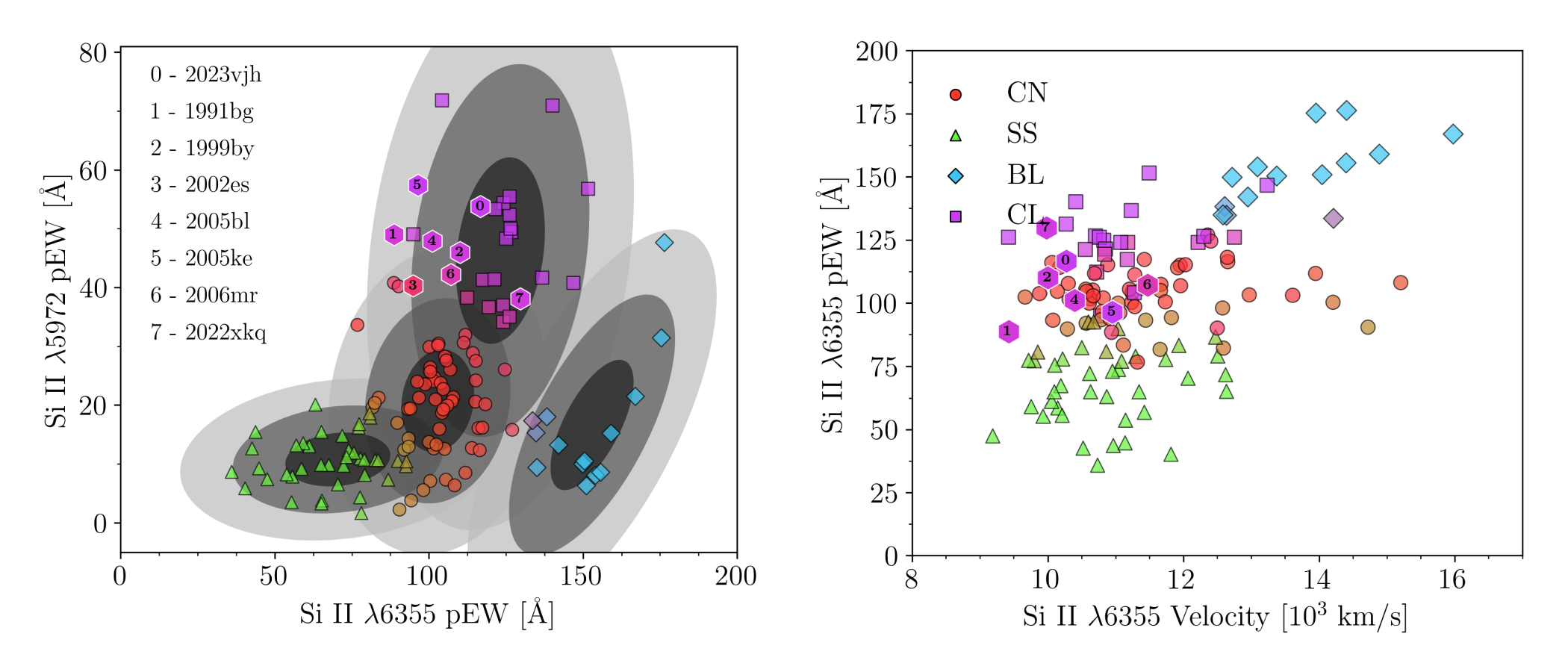}
    \caption{
\citet{2006PASP..118..560B} and  \citet{2009ApJ...699L.139W} diagrams (left and right panels respectively). For the Branch diagram we used the code developed by \citet{2020ApJ...901..154B}, formulated to determine group membership probabilistically. Different colors and shapes correspond to different classes (CN, SS, BL, CL) of SNe from the CSP-I+II \citep{2013ApJ...773...53F,2017AJ....154..211K,2019PASP..131a4002H,2019PASP..131a4001P,2024ApJ...967...20M}. In both cases SN~2023vjh is marked with the number zero. For SN~2023vjh the $ pEW_{\rm Si\,II\, 6355} = 116.7 \pm 9.4\,$\AA, $pEW_{\rm Si\,II\, 5972} = 53.8 \pm 14.1\,$\AA, and $ v_{\rm Si\,II\, 6355} = 10269 \pm 498\, \rm km\, s^{-1}$.}
    \label{fig:branch}
\end{figure*}

\begin{figure*}[htbp]
    \centering
    \begin{subfigure}{0.48\textwidth}
        \centering
        \includegraphics[width=\linewidth]{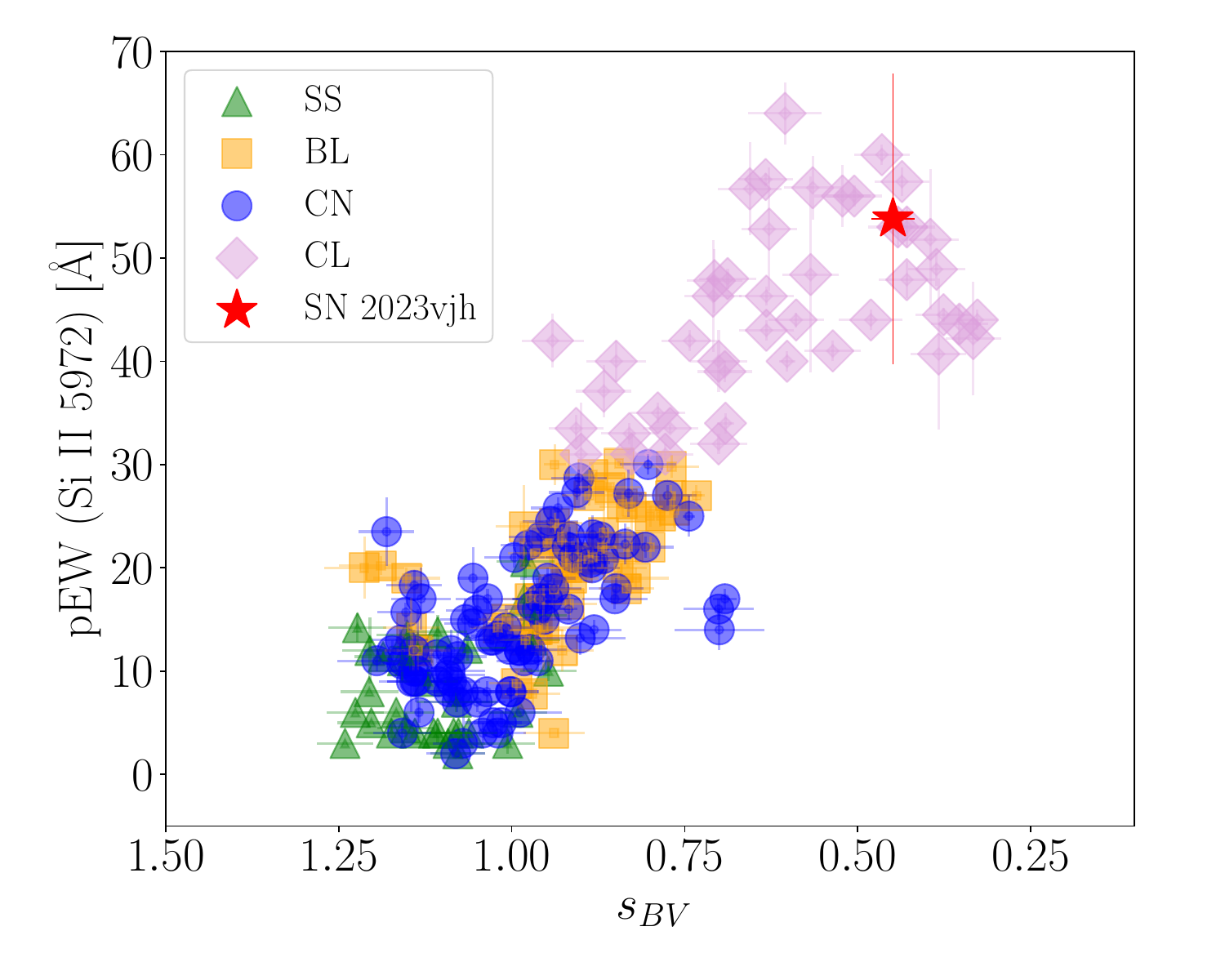}
    \end{subfigure}
    \hfill
    \begin{subfigure}{0.48\textwidth}
        \centering
        \includegraphics[width=\linewidth]{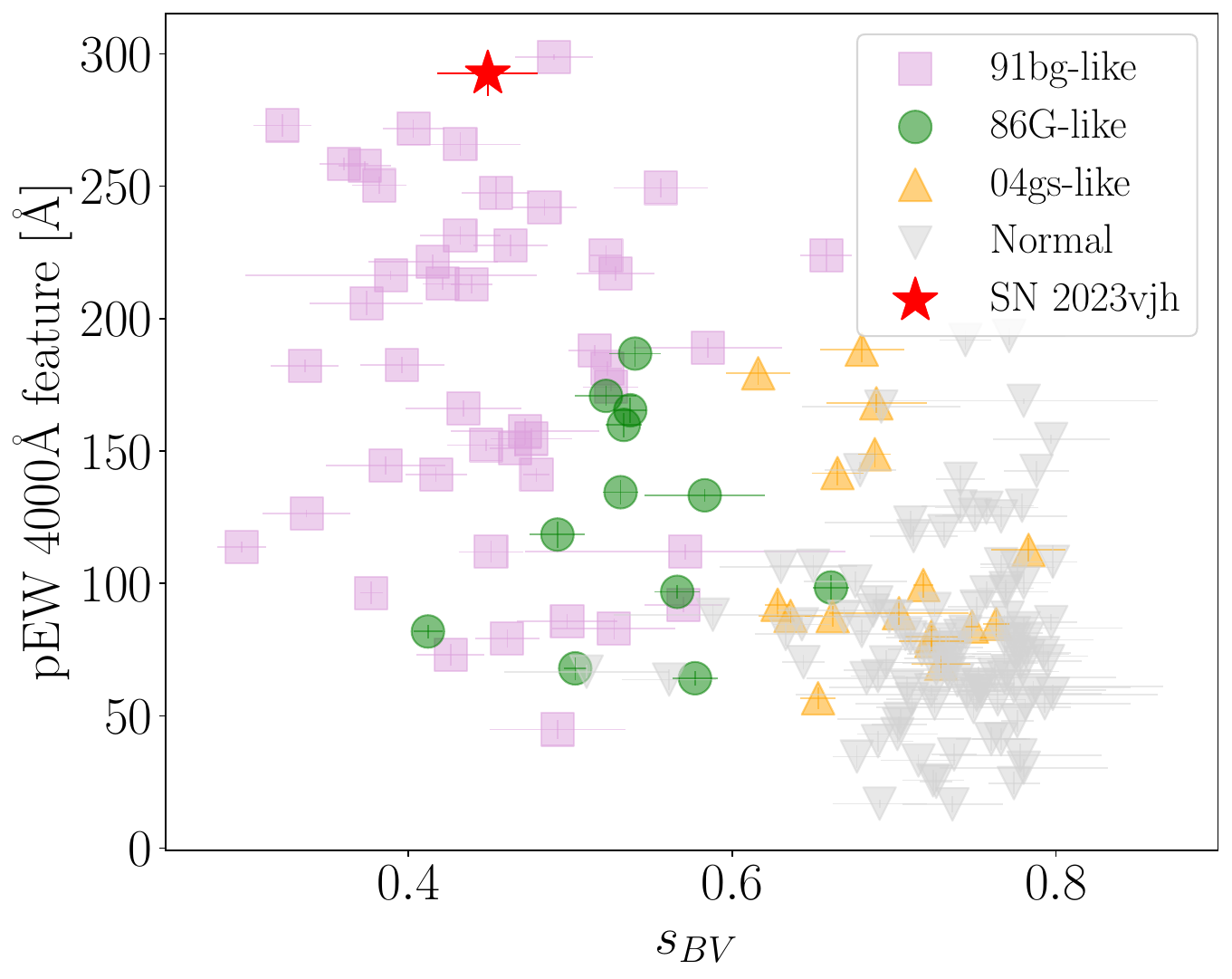}
    \end{subfigure}

    \caption{(a) The $pEW$ of \ion{Si}{II} $\lambda5972$ as a function of $s_{BV}$. The red star indicates the position of SN~2023vjh. The rest of the sources have been obtained from \citet{2024ApJ...967...20M}; (b) A $pEW$ diagram of the 4000\AA\ feature as a function of $s_{BV}$, for four SNe Ia groups, as presented in \citet{2026A&A...707A..91A}.}
    \label{fig:Si_SBV_TiII}
\end{figure*}
 Agreement of SN~2023vjh with other cool SNe Ia is also seen in Fig.~\ref{fig:dv20}, where we show the difference in the expansion velocity of \ion{Si}{II} $\lambda6355$ between maximum light and 20 days after B-band maximum, as a function of $\rm \Delta m_{15}$ for the sample of CSP-I, CSP-II, and historical SNe Ia \citep{2024ApJ...967...20M}. The different classes of SNe Ia presented in this diagram are: Core Normal (CN), Cool (CL), Shallow Silicon (SS), Broad Lined (BL). SN~2023vjh is located in the region of cool SNe Ia, which occupies the rightmost part of the plot.

In Fig.~\ref{fig:branch}, we show on the left the \citet{2006PASP..118..560B} diagram and on the right the \citet{2009ApJ...699L.139W} diagram, where different subcategories of SNe~Ia are distinguished. For the Branch diagram we used the code developed by \citet{2020ApJ...901..154B}, formulated to determine group membership probabilistically. Different colors and shapes correspond to different classes (CN, SS, BL, CL) of SNe from the CSP-I+II \citep{2013ApJ...773...53F,2017AJ....154..211K,2019PASP..131a4002H,2019PASP..131a4001P,2024ApJ...967...20M}, while the SNe 2023vjh, 1991bg, 1999by, 2002es, 2005bl, 2005ke, 2006mr, and 2022xkq \citep{2024ApJ...960...29P}  have been included indicated by a number. In both cases, SN~2023vjh occupies the region associated with CL SNe. In the Branch diagram, it lies well within the cool region, where 91bg-like SNe Ia are typically found. In the \citet{2009ApJ...699L.139W} diagram, it is located closer to the area that overlaps with the core-normal SNe Ia.
Figure \ref{fig:Si_SBV_TiII} (a) shows the $pEW$ width of \ion{Si}{II} $\lambda5972$ versus  $\rm s_{BV}$,  for the CSP-I+II, and historical SN~Ia samples \citep{2024ApJ...967...20M}. SN~2023vjh is located in the rightest portion of the locus of cool SNe.

Additionally, we examined whether SN~2023vjh falls into the “extreme cool” (eCL) subclass as defined by \citet{2013ApJ...773...53F}. Cool SNe are characterized by strong absorption around 4300 \AA, primarily due to prominent \ion{Ti}{II} features. According to \citet{2013ApJ...773...53F}, eCL SNe are those with $ pEW > 220$ \AA\ near maximum light (they call it $ pEW_{\rm Mg\,II}$ but here we will use the name $ pEW_{\rm Ti\,II}$), and they typically show a weighted-average $ pEW_{\rm Ca\,II\,IR}$ of $325 \pm 39$ \AA. We measured these $pEW$s for SN~2023vjh around maximum (using the weighted mean over $\pm$ 5 days) and obtained $ pEW_{\rm Ti\,II} = 292.5 \pm 8.3$ and $ pEW_{\rm Ca\,II\,IR} = 320.2 \pm 3.2$. Based on the Ti II criterion, SN~2023vjh therefore qualifies as an eCL SN, placing it in the same group as SN~2005bl, SN~2005ke, SN~2006bd, SN~2006mr, SN~2007ax, and SN~2009F reported in \citet{2013ApJ...773...53F}. Finally, in Fig. \ref{fig:Si_SBV_TiII} (b), using the $pEW$ of the absorption feature at ~4000 Å, we examined where SN~2023vjh is located in the pEW(4000 Å)–$s_{BV}$ diagram, in comparison with different SN groups \citep{2026A&A...707A..91A}. SN~2023vjh occupies the upper region of the 91bg-like SNe.

\end{appendix}

\end{document}